\documentclass[%
 reprint,
 amsmath,amssymb,
 aps,
floatfix,
]{revtex4-1}

\usepackage{graphicx}
\usepackage{dcolumn}
\usepackage{bm}
\usepackage{scalerel}
\usepackage{textcomp}
\usepackage{amsmath}
\usepackage{physics}
\usepackage[caption=false]{subfig}
\usepackage{booktabs}

\usepackage{xcolor}
\definecolor{darkblue}{RGB}{0,0,150}
\definecolor{nightblue}{RGB}{0,0,100}
\usepackage[
bookmarksnumbered,
pdfpagelabels=false,
colorlinks,
linkcolor=nightblue,
citecolor=nightblue,
filecolor=black,
urlcolor=darkblue,
breaklinks=true
]{hyperref}
\begin{document}

\preprint{APS/123-QED}

\title{Induced Half-Metallicity and Gapless Chiral Topological \\ Superconductivity in the CrI$_3$-Pb Interface}

\author{Gilad Margalit}
\author{Binghai Yan}
\author{Yuval Oreg}
 \affiliation{Department of Condensed Matter, Weizmann Institute of Science.}

\date{\today}

\begin{abstract}
We study a two-dimensional heterostructure comprised of a monolayer of the magnetic insulator chromium triiodide (CrI$_3$) on a superconducting lead (Pb) substrate. Through first-principles computation and a tight-binding model, we demonstrate that charge transfer from the Pb substrate dopes the CrI$_3$ into an effective half-metal, allowing for the onset of a gapless topological superconductivity phase via the proximity effect. This phase, in which there exists a superconducting gap only in part of the Fermi surface, is shown to occur generically in 2D half-metal-superconductor heterostructures which lack two-fold in-plane rotational symmetry. However, a sufficiently large proximity-induced pairing amplitude can bring such a system into a fully-gapped topological superconducting phase. As such, these results are expected to better define the optimal 2D component materials for future proposed TSC heterostructures.
\end{abstract}

\maketitle

\section{Introduction}
\label{Introduction}

Topological superconductivity (TSC) remains one of the most intriguing concepts in modern condensed matter physics, with useful properties such as robust, low-energy edge modes and quasiparticles that exhibit non-Abelian statistics \cite{Nayak2008}. In recent years, several experiments have displayed evidence of a quasi-1D topological phase in semiconductors such as InAs and InSb in proximity with s-wave superconductors (for a review of nanowire structures see \cite{Lutchyn2018}), Fe ad-atoms on Pb \cite{Yazdani2014}, full-shell wires \cite{Vaitiekenas2018}, and Josephson junctions \cite{Yacoby2019, Fornieri2019}. However, topological superconductivity in 2D has proven to be more elusive, despite the recent discovery of dozens of 2D materials which could form the basis for layered heterostructures \cite{Heine2014}. Successfully producing this novel topological phase of matter could potentially pave the way for nanotechnology applications and quantum computation.

One promising theoretical proposal for realizing a chiral TSC phase involves a half-metal (HM), a metal with only one spin component at the Fermi energy due to a strong exchange field, in proximity with a s-wave superconductor substrate (described in 1D by \cite{Brouwer2011} and in 2D by \cite{Sau2010,SCZhang2011,Hao2016}). The half-metal may be a 2D ferromagnetic layer, and indeed an experiment using an Fe monolayer for this purpose has demonstrated a convincing chiral edge signature~\cite{Wiesendanger2019}. Strong Rashba spin-orbit coupling (SOC) in the substrate, or another source of effective spin-flip hopping between the HM and superconductor, is additionally necessary, as we show in Sec. \ref{Theory}.

Our goal is to design a 2D heterostructure of this type, making use of the specific band structures of currently-realizable materials. We chose lead (Pb), a material with large SOC interactions, as the substrate, and monolayer CrI$_3$ (either exfoliated \cite{Baral2019} or grown by molecular beam epitaxy \cite{Chen2019,Li2019}) as the half-metal. Although CrI$_3$ is a ferromagnetic insulator in the monolayer limit \cite{Huang2017}, we demonstrate with a first-principles calculation that half-metallicity can be induced via charge transfer from the Pb side to the CrI$_3$ side (Sec. \ref{DFT}).

The presence of topological edge modes requires an odd number of effectively spinless bands at the Fermi energy in this scheme \cite{Sato2010}. Following the charge transfer, our computation shows that a single, spin-polarized band from the CrI$_3$ intersects the Fermi energy.

Since the first-principles density functional theory (DFT) simulation does not support superconductivity, we examine the effects of superconductivity in a two-step process. First, we create a tight-binding model that captures the salient features of the first-principles results (Sec. \ref{TightBindingModel}). Then, we extend this model to include the superconductivity of the Pb substrate via the addition of a phenomenological pair potential $\Delta$ using the Bogoliubov-de-Gennes (BdG) formalism. (Sec. \ref{SC}).

When we add superconductivity in this way, we observe that, rather than forming a fully-gapped TSC phase, this system will induce topological gaps only in small regions of its Fermi surface. This realizes a gapless TSC phase similar to those described in \cite{Wong2013,Baum2015,Daido2017,Kamenev2018}, characterized by topological edge modes coexisting with bulk states.

The phase arises due to a property shared by the majority of magnetic monolayers discovered so far - an absence of C$_2$ symmetry (symmetry under 180$^\circ$ in-plane rotation). Without this symmetry, electrons with wave vector $k$ do not have a partner state at the same energy at $-k$ with which to form a superconducting pair (see Figs. \ref{BdGsymm} and \ref{fermi}). Since this symmetry is rare in known monolayers, these results are expected to be applicable to a broad class of heterostructures. 

\section{Theory}
\label{Theory}

Here, we discuss the theoretical framework in which we expect the interface between Pb and CrI$_3$ to allow for a chiral TSC phase. We define our system in terms of the creation operators $c_{\textbf{k}s}^\dagger$, with momentum $\textbf{k} = (k_x, k_y)$ and spin $s$, of a single hybrid band that crosses the Fermi energy, composed of states lying primarily within the interface between the two materials.

The Hamiltonian of this system can be described by:

\begin{eqnarray}
\label{Hamiltonian}
\begin{aligned}
H &= H_\text{0} + H_\text{SC} + H_\text{XC} + H_\text{SOC}; \\
H_\text{0} &= \sum_{\textbf{k},s}{(\epsilon_{\textbf{k}}-\mu) c_{\textbf{k}s}^\dagger c_{\textbf{k}s}} \\
H_\text{SC} &= \Delta \sum_{\textbf{k}}{(c_{\textbf{k}\uparrow}^\dagger c_{-\textbf{k}\downarrow}^\dagger + c_{\textbf{k}\uparrow}c_{-\textbf{k}\downarrow})} \\
H_\text{XC} &= J \sum_{\textbf{k},s}{c_{\textbf{k}s}^\dagger \sigma_{s s'}^z c_{\textbf{k}s'}} \\
H_\text{SOC} &= \alpha \sum_{\textbf{k},\sigma}{c_{\textbf{k}s}^\dagger(k_x \sigma_{s s'}^y - k_y \sigma_{s s'}^x) c_{\textbf{k}s'}}.
\end{aligned}
\end{eqnarray}

Here, the surface of the superconductor and the 2D half-metal monolayer coupled to it both contribute terms to the hybrid system Hamiltonian via proximity. $H_0$ describes a single effective spinful band with dispersion $\epsilon_{\textbf{k}}$ and chemical potential $\mu$. $\Delta$ is an effective s-wave pair potential at the surface, which we take to be on-site (constant in momentum-space). The half-metal contains a large exchange potential $J$, causing the band to become strongly spin-polarized ($\sigma^i$ represent Pauli matrices in the spin degree of freedom), where the spin quantization axis is perpendicular to the surface. A Rashba-type spin-orbit interaction \cite{Rashba} with magnitude $\alpha$ is additionally allowed since the joining of the two surfaces breaks the out-of-plane component of inversion symmetry ($M_z$).

Chiral topological superconductivity requires that opposite-momentum states of a single band at the Fermi energy be paired by $\Delta$; the above Hamiltonian accomplishes this by gapping out one of the spin bands through the half-metal's large exchange interaction. Though the remaining band contains only spin-up states, SOC interactions ($\alpha$) cause them to take on a small in-plane component, of order $\alpha / J$. Since, for this particular type of spin texture, states at opposite momenta have opposite in-plane spins, they can be coupled by the $\Delta$ term in Eq. \ref{Hamiltonian}. The resulting effective p-wave pair potential scales as $\tilde{\Delta} \sim \frac{\alpha}{J}\Delta$.

This framework characterizes the important features of the Pb-CrI$_3$ interface system: a large exchange potential, which originates in the CrI$_3$ monolayer, SOC from both materials, and superconductivity from the Pb substrate. Crucially, we find via DFT (see Sec. \ref{DFT}) that charge transfer from the Pb to the CrI$_3$ raises the Fermi energy, causing the CrI$_3$ to change from a ferromagnetic insulator to an effective half-metal. Formally, the first conducting band above the gap now crosses the Fermi energy.

\begin{figure}
     \subfloat[]{\includegraphics[width=0.23\textwidth]{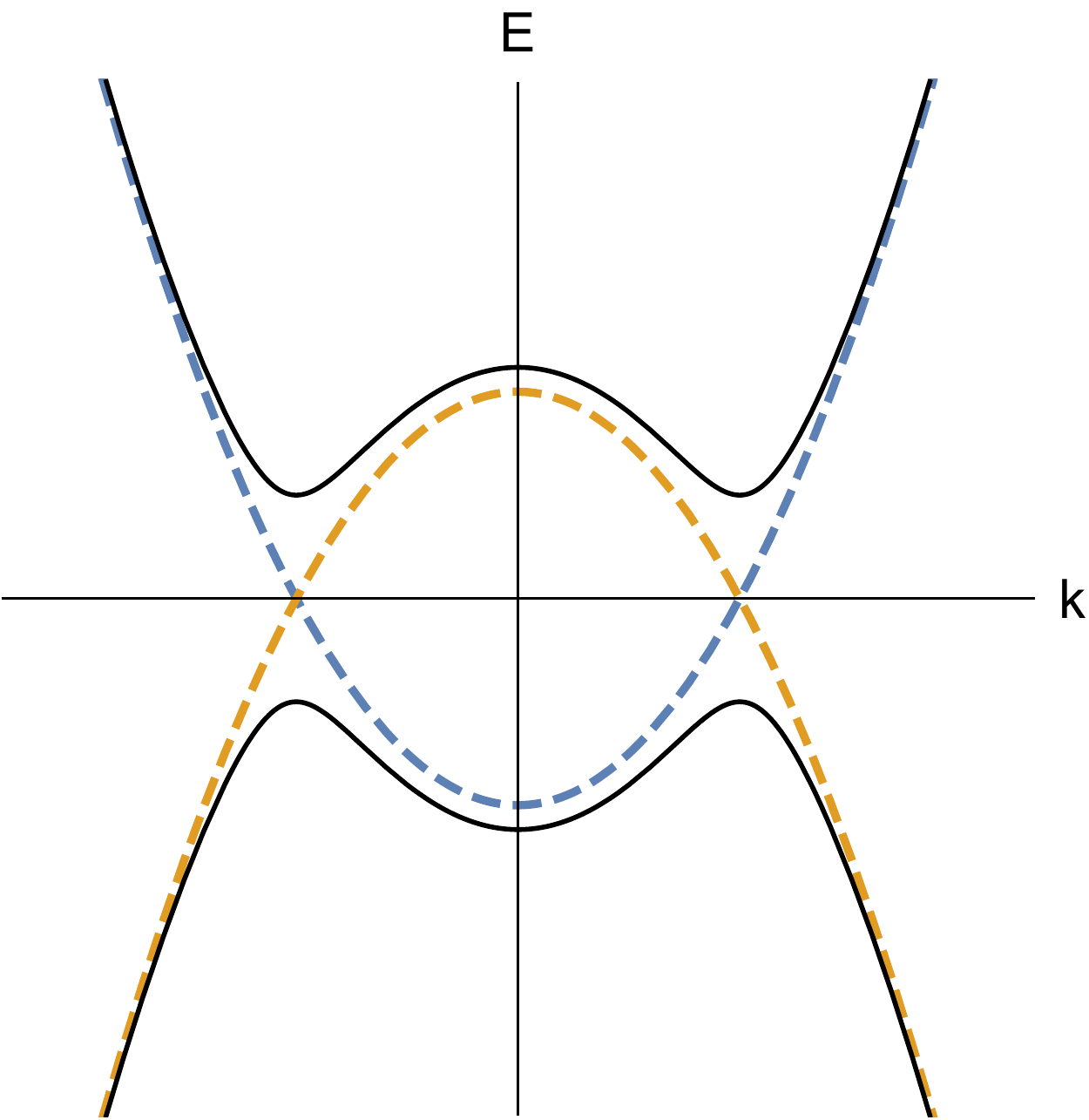}
         \label{symmBdG}}
     \subfloat[]{\includegraphics[width=0.23\textwidth]{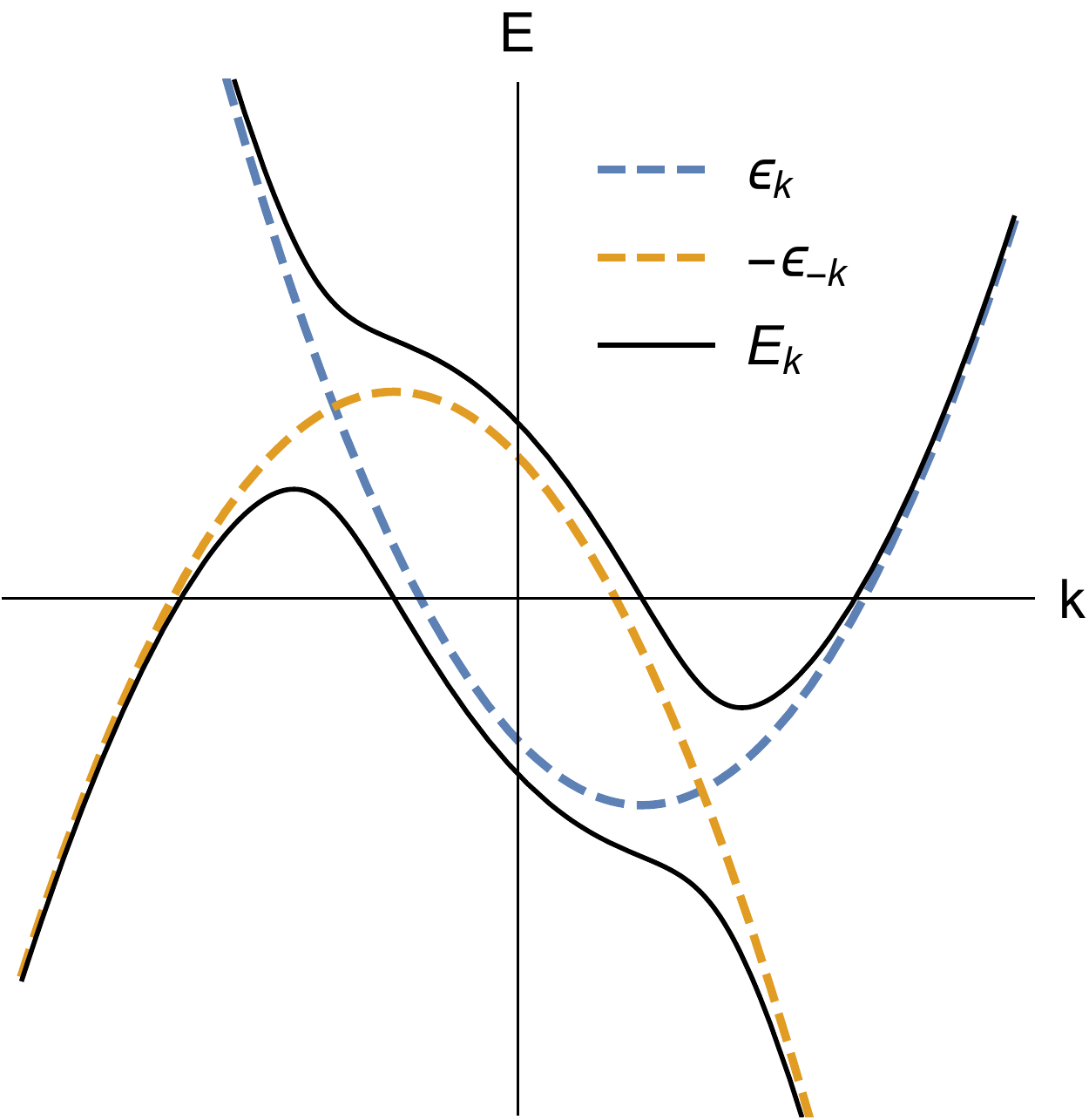}
         \label{asymmBdG}}
         
        \caption{Schematics of inversion-symmetric (a) and inversion-asymmetric (b) bands. We plot the bare dispersions $\epsilon_k$ and $-\epsilon_{-k}$ (dashed lines) and their corresponding dispersion $E_k$ when a mean-field pair potential is added (solid lines). While the symmetric case guarantees that there are no unpaired bulk states at the Fermi energy, this is not guaranteed for the asymmetric case. Though depicted here in 1D, this latter case generalizes straightforwardly to 2D, in which regions of the Fermi surface are occupied by bulk states in the gapless TSC phase.}
        \label{BdGsymm}
\end{figure}

As we will see in Sec.~\ref{DFT}, due to symmetry breaking when the Pb is brought into contact with the CrI$_3$, the dispersion becomes inversion-asymmetric. In general, when $\epsilon_{\textbf{k}} \neq \epsilon_{\textbf{-k}}$, then the superconducting gap may be misaligned with the Fermi energy, leading to gapless superconductivity. We illustrate this with a simple parabolic dispersion in Fig. \ref{asymmBdG}. Though such systems are not fully gapped, they may still exhibit nontrivial topology, as we show in Sec. \ref{results}. In this case, topological edge modes cease to be fully protected from disorder (since there are nearby bulk states at the same energy for them to scatter to), but they do not disappear altogether.

\section{First-Principles Calculation}
\label{DFT}

In this section, we describe the first-principles calculation that is used to derive the band structure of the hybrid system, which will be used as the basis for a tight-binding model in proximity to superconductivity. Our simulations made use of density functional theory (DFT), employing the Vienna \textit{Ab-initio} Simulation Package (VASP)\cite{Kresse1993, Kresse1999} with generalized gradient approximation (GGA) functionals \cite{GGA}.

\begin{figure}
     \subfloat[]{\includegraphics[width=0.45\textwidth]{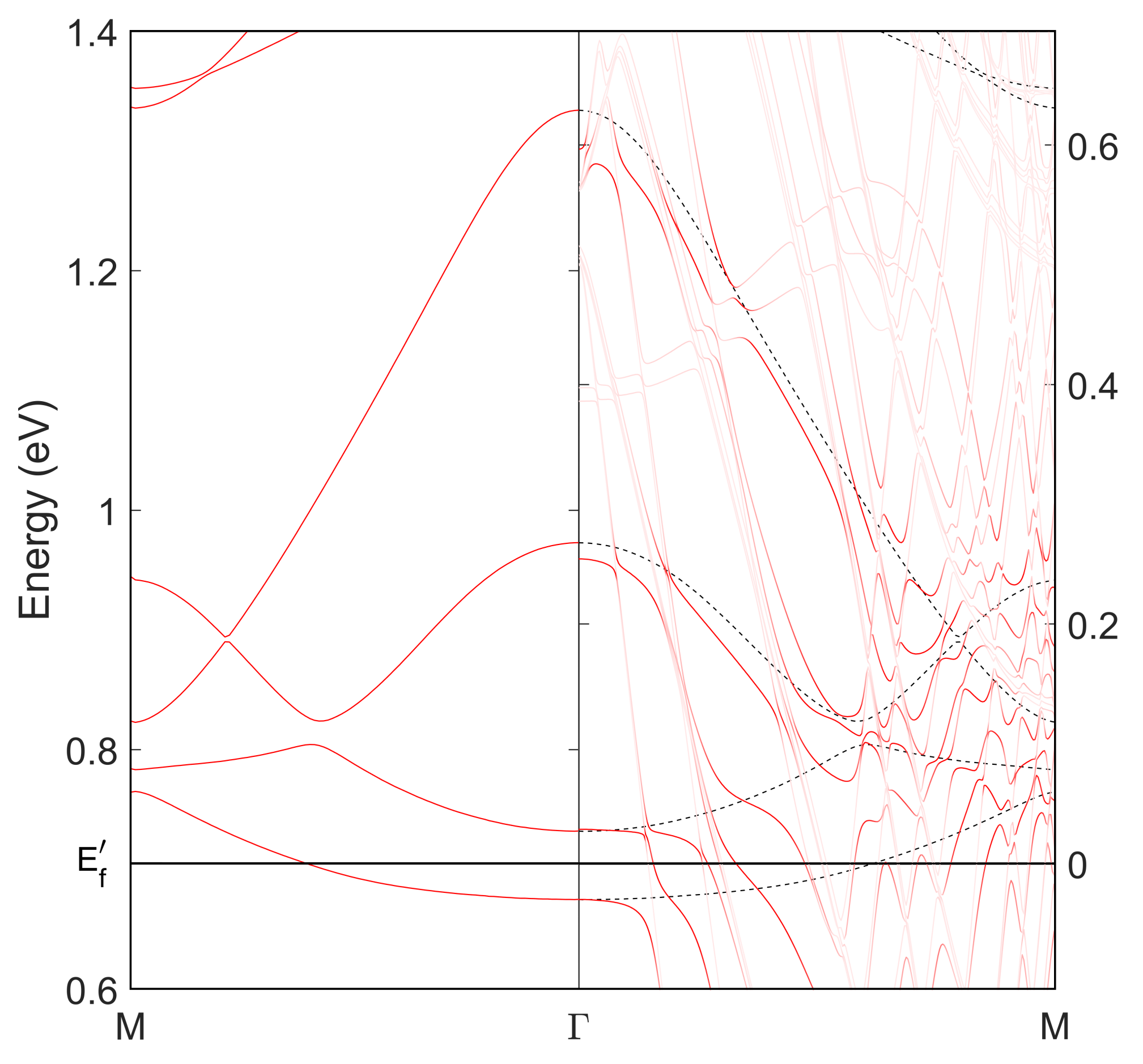}
         \label{bandCompare}}
         
     \subfloat[]{\includegraphics[width=0.48\textwidth]{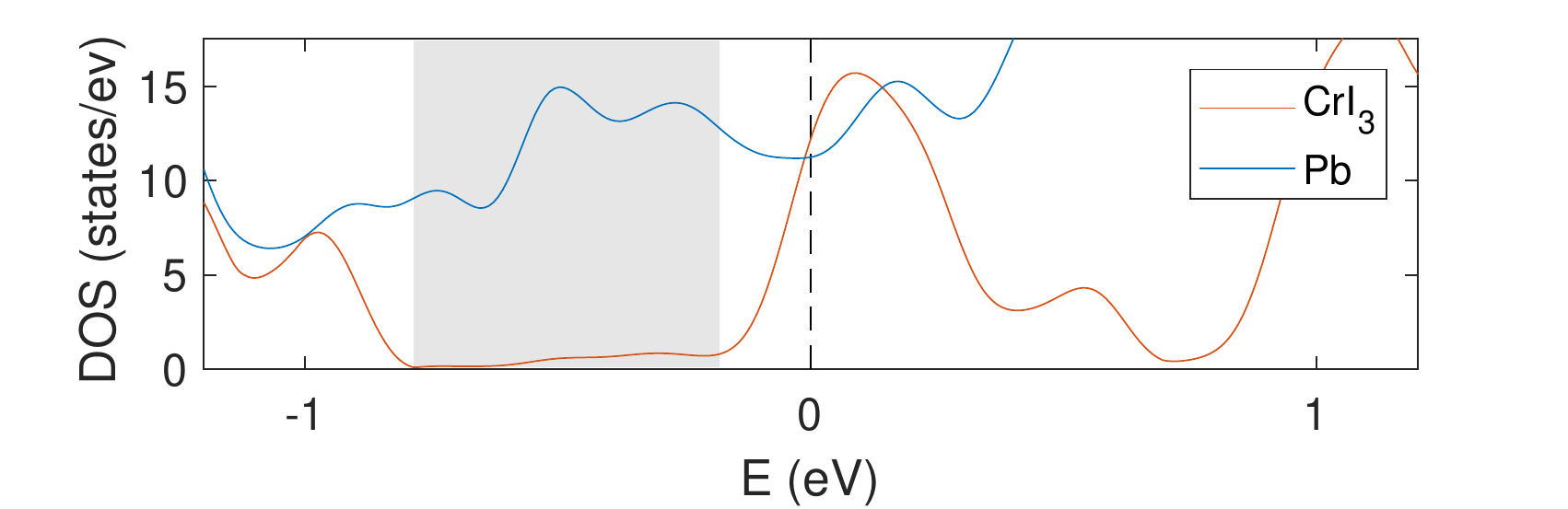}
         \label{interfaceDOS}}
        \caption{(a) Band structures of the CrI$_3$ monolayer and the CrI$_3$-Pb interface. Left: band structure of isolated CrI$_3$ monolayer.  Right: band structure of the interface when the CrI$_3$ is in contact with the Pb substrate. The Fermi energy is set to zero. Line intensity in the right panel indicates projection to the interface atoms (CrI$_3$ and the top Pb atomic layer). Among the interface states, the band dispersion of CrI$_3$ (dashed lines) can be discerned. We see that, due to the doping effect by the charge transfer, the lowest conduction band of CrI$_3$ is shifted to the Fermi level ($E_f^{\prime}$) at the interface. (b) Density of states (DOS) projected to the CrI$_3$ layer (red) and the Pb substrate (blue) in the interface system. Due to electron doping, the Fermi energy shifts from inside the $CrI_3$ gap (shaded region with small DOS) in the isolated case to crossing the lowest conduction band (with large DOS) in the interface case.}
        \label{bands}
\end{figure}

\begin{figure}
     \subfloat[]{\includegraphics[width=0.23\textwidth]{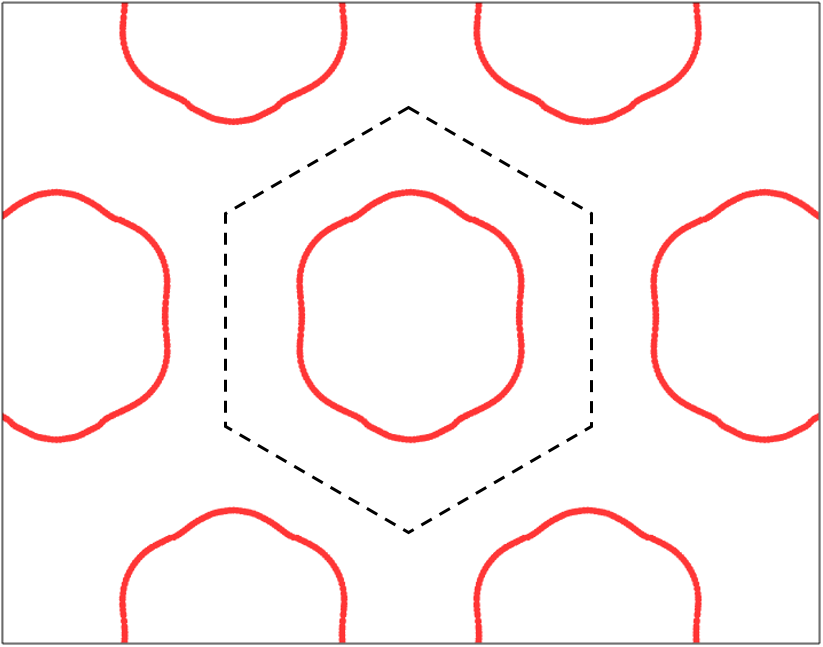}
         \label{cri3fermi}}
     \subfloat[]{\includegraphics[width=0.23\textwidth]{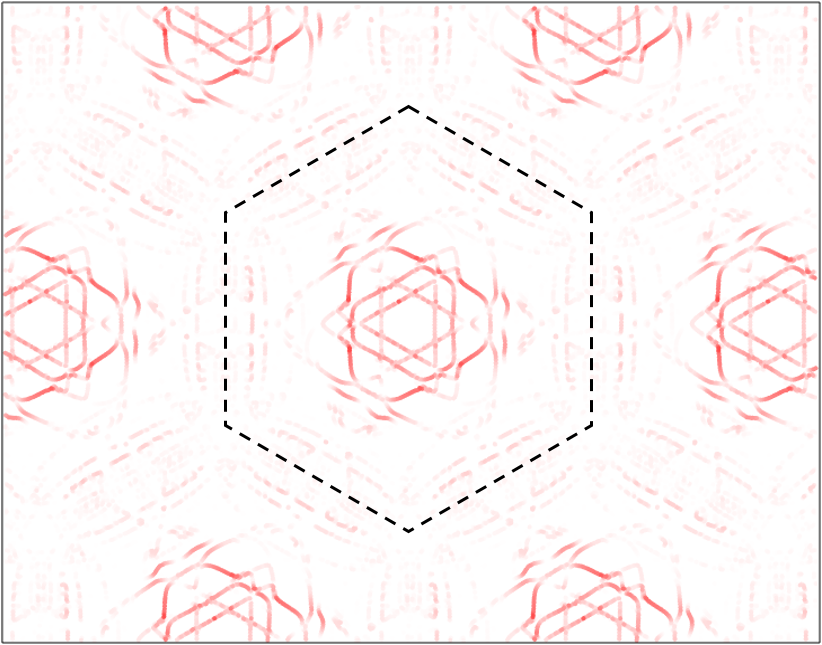}
         \label{interfacefermi}}
        \caption{(a) Fermi surface of an isolated CrI$_3$ monolayer in 2D momentum space at the same energy above the bottom of the band as in the heterostructure case. (b) Fermi surface of the interface. Intensity indicates the projection amplitude to the Cr and I atoms; the largest triangular curve in each Brillouin zone is the dominant feature and can be viewed as the curve in (a) distorted by hybridization with the Pb.}
        \label{fermi}
\end{figure}

\begin{figure}
     \subfloat[]{\includegraphics[width=0.23\textwidth]{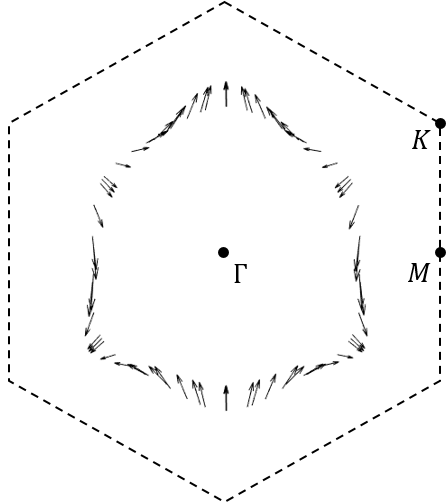}
         \label{cri3spinfermi}}
     \subfloat[]{\includegraphics[width=0.23\textwidth]{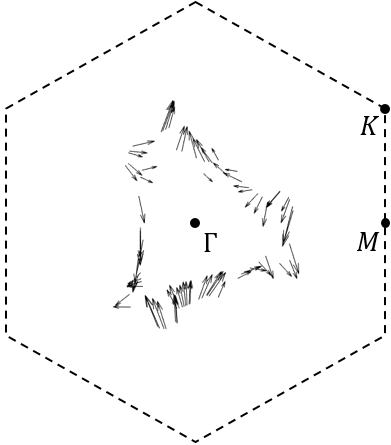}
         \label{interfacespinfermi}}
         
     \subfloat[]{\includegraphics[width=0.23\textwidth]{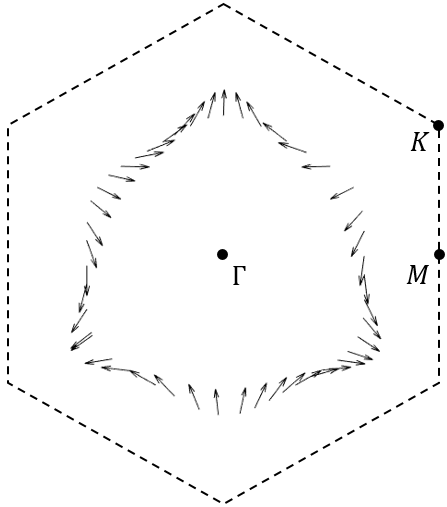}
         \label{twist}}
     \subfloat[]{\includegraphics[width=0.225\textwidth]{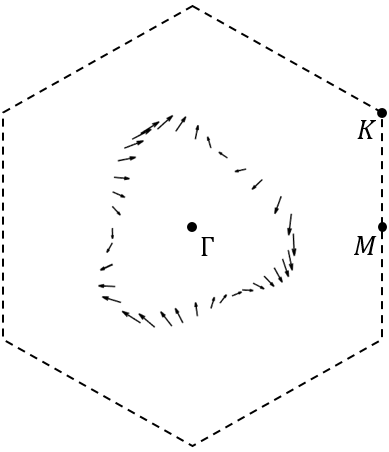}\llap{\includegraphics[height=2cm]{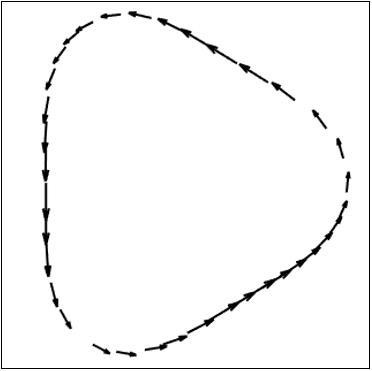}}
         \label{twistAsym}}
        \caption{(a) In-plane spin vectors of isolated CrI$_3$ in the first Brillouin zone, computed with DFT. The Fermi level is taken to be the same energy above the bottom of the band as in the interface case so as to provide a basis of comparison; the true Fermi level of isolated CrI$_3$ lies within an insulating gap and would thus be featureless. (b) In-plane spin vectors of the Pb-CrI$_3$ interface states. For clarity, only interface states strongly associated with the CrI$_3$ layer are plotted (see Sec. \ref{DFT} for details). The omitted spins correspond to the Pb bands.
        (c) Spin texture obtained from a tight-binding model (Section \ref{TightBindingModel}). Tight-binding coefficients were chosen to match qualitatively the CrI$_3$ spin texture shown in (a) (reproducing the shape and size of the band and the magnitudes and directions of the spin vectors). (d) Modeled in-plane spin texture with coefficients chosen to qualitatively match the interface texture in (b). It is a superposition of two spin textures, the in-plane texture of isolated CrI$_3$, which winds twice when traced around the Fermi pocket,and the desired Rashba spin texture (inset), which winds once.}
        \label{spinfermi}
\end{figure}

A 2-by-2 super-cell of the Pb (111) plane has a closely-matching lattice constant (1.4\% larger) to the primitive unit cell of CrI$_3$. Thus, we constructed the interface from a CrI$_3$ monolayer on top of  the (111) surface of a slab of~Pb, assuming that the monolayer would expand to match the lattice constant of the substrate. We carefully tested 13 to 31 atomic layers of Pb and obtained qualitatively the same results for the band structure. The DFT results shown in Figs. \ref{bands}-\ref{spinfermi} are based on a 13-layer-thick slab.

\begin{figure}
     \subfloat[]{\includegraphics[width=0.45\textwidth]{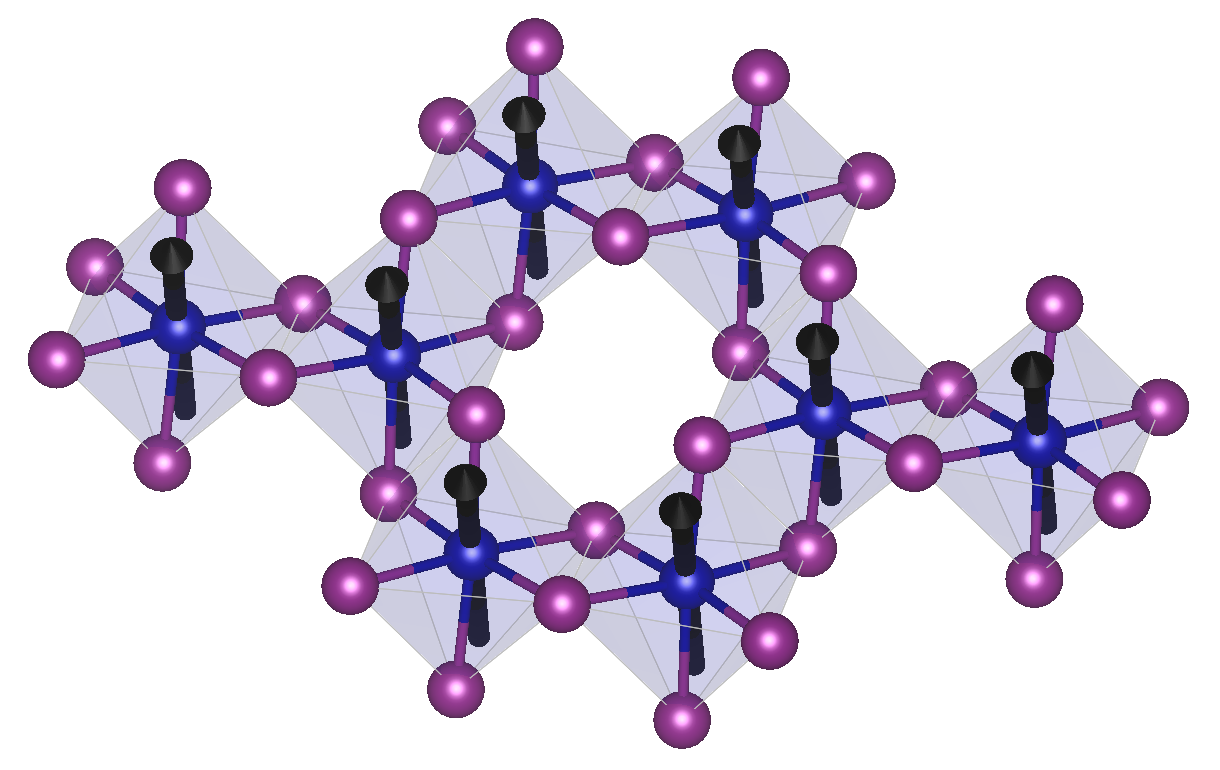}
        \label{CrI3_fig}}
        
     \subfloat[]{\includegraphics[width=0.21\textwidth]{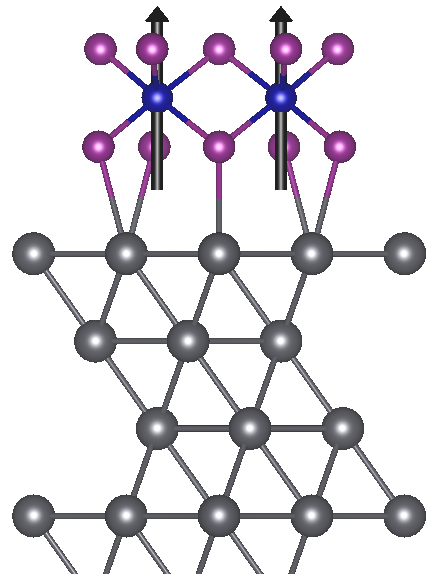}
        \label{interfaceVesta}}
     \subfloat[]{\includegraphics[width=0.24\textwidth]{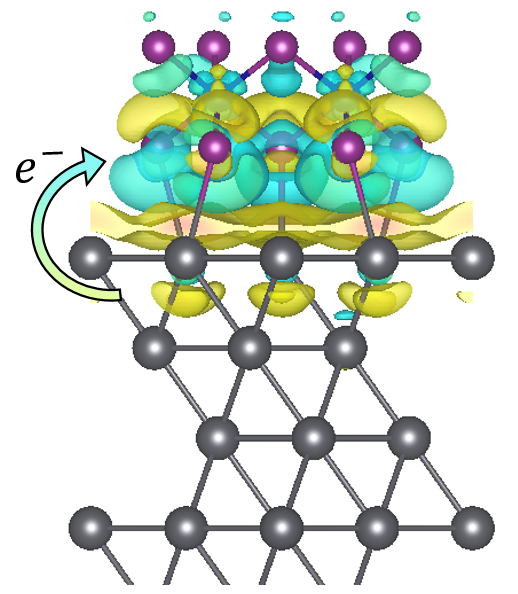}
        \label{chargediff}}
        \caption{(a) CrI$_3$ lattice with each Chromium atom (blue) bonded to six Iodine atoms (purple) in an octahedral configuration, forming a 2D honeycomb. Black arrows indicate the magnetic moments of the Cr atoms. (b) A unit cell of CrI$_3$ bonded to bulk Pb (gray). This diagram shows the top 4 lead layers input to DFT. (c) Electron density difference between the cases when the materials are in isolation and when they are placed in contact, computed with DFT. Cyan (yellow) indicates regions where the density has increased (decreased) after the Pb and CrI$_3$ are brought into contact. We see an overall movement of electrons from the Pb into the CrI$_3$, explaining the increase in Fermi energy that transforms the CrI$_3$ layer from a ferromagnetic insulator into a half-metal.}
        \label{vesta}
\end{figure}

 All band structure calculations were preceded by ionic relaxation, in which the surface monolayer and substrate bonded in the manner shown in Fig. \ref{interfaceVesta}. In this step, we optimized the position of the Cr and I atoms while fixing the Pb atoms in place to simulate that a monolayer bonds to a half-infinite Pb substrate. The calculations also included SOC interactions in all materials.

In the band structure, contact with the Pb surface causes the CrI$_3$ bands to hybridize with the superimposed Pb bands, as shown on the right-hand side of Fig. \ref{bandCompare}. Despite the strong hybridization, we can still identify the dispersion of CrI$_3$ conduction bands from the background of Pb states.

A salient feature is that the lowest CrI$_3$ conduction band becomes partially doped by electrons transferred from the metal substrate, raising the Fermi energy. Charge density maps suggest that electrons from the Pb are drawn to the surface when the materials are coupled (Fig. \ref{chargediff}) due to a difference in the work function between the materials. As shown in Fig. \ref{interfaceDOS}, even though a monolayer of CrI$_3$ in isolation is an insulating ferromagnet, the Fermi energy of the interface system lies above the CrI$_3$ insulating gap (shaded region). This results in a single conducting pocket for CrI$_3$.

Changing the number of Pb layers does not change this general conclusion; the Fermi energy varied by approximately $\pm 10$ meV between different Pb slabs relative to the CrI$_3$ bands, but was an average of about 30 meV above the bottom of the lowest monolayer band. In this way, we have shown that charge transfer can in principle be employed to achieve the proposed half-metal-superconductor hybrid system TSC without an intrinsic half-metal.

Crucially, the distortion to the Fermi surface from coupling with the Pb also reduces the symmetry, as shown in Fig. \ref{fermi} (to provide a basis of comparison, Fig. \ref{cri3fermi} shows the Fermi surface at the same energy above the bottom of its band as in the interface case). This is due to the fact that CrI$_3$ in isolation has inversion symmetry, but this is broken by the presence of Pb atoms in only the $-z$ half-space.

Specifically, the CrI$_3$ monolayer has $M_yM_zT$ symmetry, where $M_i$ is a mirror symmetry in the $x_i$ direction about a hexagon center and $T$ is time-reversal symmetry. This symmetry, together with $M_xT$ symmetry, makes CrI$_3$ inversion-symmetric, and in particular, $M_yM_zT$ enforces the relation $\varepsilon_n(k_x,k_y) = \varepsilon_n(-k_x,k_y)$. The presence of the Pb substrate breaks $M_yM_zT$, and thus when inversion overall is lost, so is inversion in the plane.

Such in-plane inversion-breaking is a generic feature of 2D magnetic heterostructures where one of the layers lacks two-fold in-plane rotational symmetry ($C_2$). This is because the only symmetries that protect $\varepsilon_n(k) = \varepsilon_n(-k)$ (in-plane inversion symmetry) are time-reversal, inversion(including out-of-plane), and $C_2$. However, the magnetic moment of the half-metal breaks time-reversal, and the layering along the $z$-axis breaks inversion. As a consequence, in the absence of $C_2$, not all bulk states at the Fermi energy can pair when a superconducting pair potential is introduced to the system. $C_2$ is not present in monolayer CrI$_3$, and as we will see in Sec. \ref{results}, this is directly responsible for the gapless TSC phase.

Figures \ref{cri3spinfermi} and \ref{interfacespinfermi} show the in-plane spin texture of isolated CrI$_3$ and the interface, respectively, calculated using DFT. Again, the Fermi energy of the isolated case is shifted to the same energy above the bottom of the band as in the interface case. For clarity, in the interface case, we only plot states with a projection to the Cr and I atoms greater than 50\%, then impose an additional cutoff in in-plane spin magnitude that removes the smallest 40\% of vectors out of those remaining. In both cases, the in-plane spin components are approximately 2 orders of magnitude smaller than the out-of-plane component, which is spin-up (not plotted). The isolated CrI$_3$ spin texture is inversion-symmetric, as expected, causing opposite-momentum states to have the same spin vectors.

Since pairing interactions couple opposite spin components of opposite-momentum states, we require an additional Rashba term. This term is allowed due to inversion breaking in the interface system. However, the spin texture in \ref{interfacespinfermi} is largely dominated by the inversion-preserving texture, as can be seen by its winding number of -2 rather than 1. This small magnitude of Rashba interactions will ultimately limit the size of the effective superconducting gap.

\section{Tight-Binding Model}
\label{TightBindingModel}

Here, we form a tight-binding model that reflects the band structure of the heterostructure. States in the relevant bands are tightly-bound to surface Cr atoms, and thus it suffices to consider on-site and nearest-neighbor interactions only, where each Cr atom represents a site. There are two sites per unit cell, and each site hosts two spins (up and down) and two orbitals ($d_{x^2-y^2}$ and $d_{z^2}$), for a total of eight bands. Only the lowest-energy of these eight bands crosses the Fermi energy, so free parameters that do not vanish due to symmetry are chosen so as to match the DFT result for the lowest-energy band. For details, see Appendix \ref{tb}.

This model can recreate the spin textures obtained through DFT for isolated CrI$_3$ as well as the full interface system, as can be seen by comparing the DFT results (Figs. \ref{cri3spinfermi} and \ref{interfacespinfermi}) to the modeled textures (Figs. \ref{twist} and \ref{twistAsym}). The parameters resulting in Fig. \ref{twistAsym} are detailed in Table~\ref{tab:params} (in Appendix \ref{tb}) and used throughout the remainder of this work.

\subsection{Model with Superconductivity}
\label{SC}

We add superconductivity to this model by including a phenomenological pair potential $\Delta$ using the BdG formalism. We assume that the Pb contributes an s-wave pairing potential that affects only opposite spins on the same site and in the same orbital (in Appendix \ref{triplet}, we extend this model to allow for the possibility of same-spin pairing interactions arising at the surface of the SC substrate). This results in the following 16-band Hamiltonian:

\begin{eqnarray}
\label{BdG}
H_{BdG} = 
\begin{pmatrix}
H_{ijk}(k)\lambda^i \sigma^j \omega^k& i\Delta \lambda^0 \sigma^2 \omega^0 \\
-i\Delta \lambda^0 \sigma^2 \omega^0& -(H_{ijk}(-k)\lambda^i \sigma^j \omega^k)^*
\end{pmatrix}.
\end{eqnarray}

Here, $\lambda^i$, $\sigma^j$, and $\omega^k$ are Pauli matrices in the sublattice, spin, and orbital degrees of freedom, respectively. The 8-band Hamiltonian in the absence of superconductivity is parametrized by the momentum-dependent terms $H_{ijk}(k)$ of our tight-binding model.

\section{Results}
\label{results}
\begin{figure}
     \subfloat[]{\includegraphics[width=0.45\textwidth]{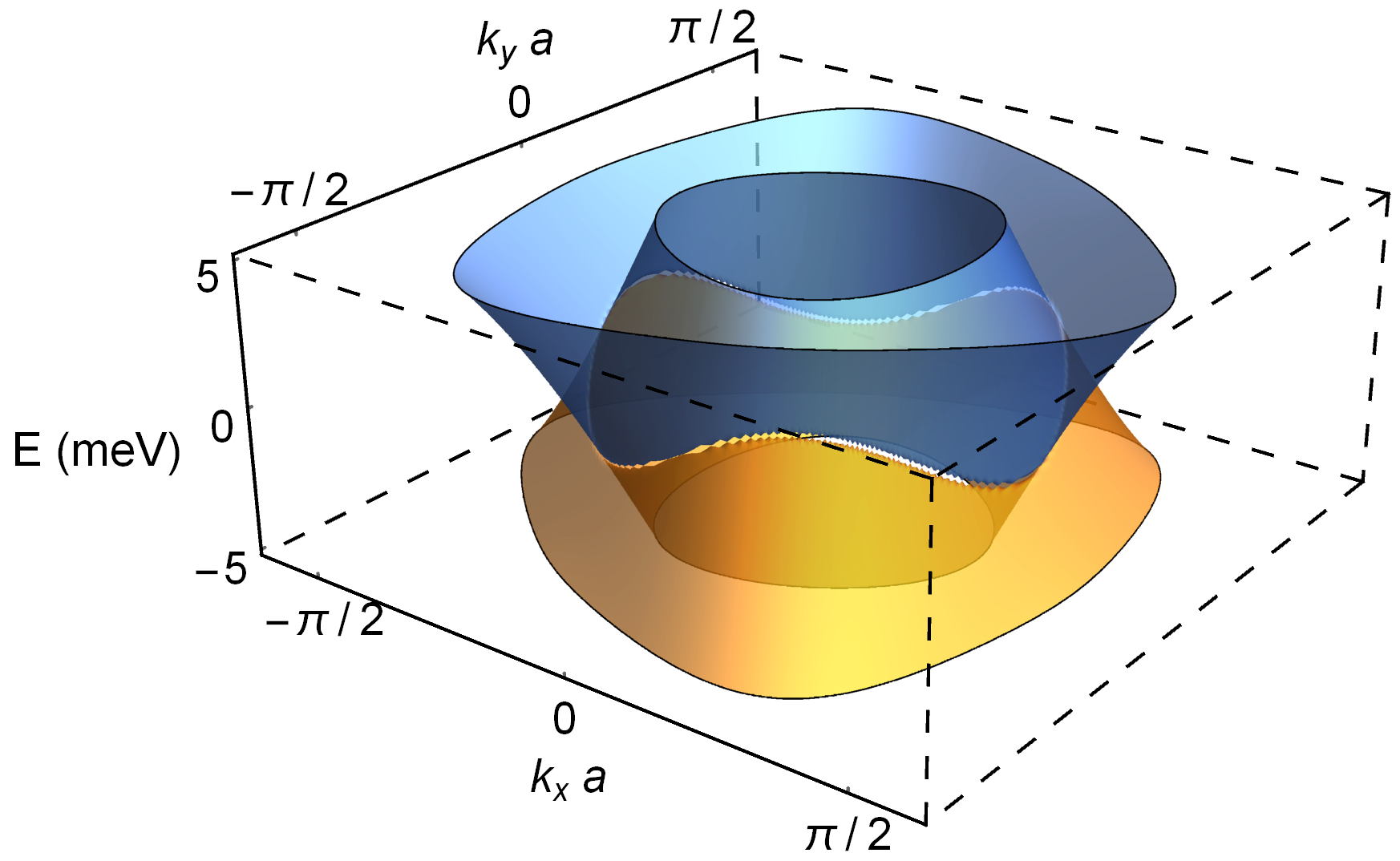}
         \label{fermi3d}}
         
     \subfloat[]{\includegraphics[width=0.3\textwidth]{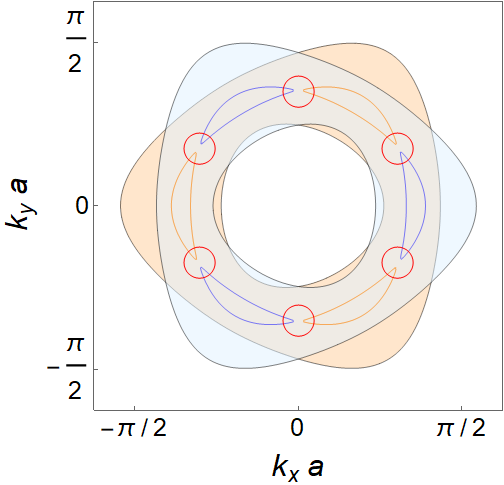}
         \label{fermicontour}}
        \caption{(a) Spectrum in momentum space near the Fermi energy (E = 0) of the lowest band and its particle-hole conjugate band at finite $\Delta$. Their intersection oscillates above and below the Fermi energy due to the triangular band structure. (b) Contour plot of the same bands in (a) at zero energy, with overlays of the curve at $\pm$5 meV (filled by light orange and light blue). At E = 0, there are 6 metallic Fermi curves (blue and orange), with gaps due to $\Delta$ only at the small hybridization regions between them (red circles).}
        \label{fig:threegraphs}
\end{figure}

We apply our model to indicate the behavior of the interface system when the Pb is made to superconduct. As shown in the modeled Fermi surface at low energies (Fig.~\ref{fermi3d}), when we add superconductivity to the case where inversion symmetry is broken by the Pb substrate, the intersection of the lowest band and its particle-hole conjugate band oscillates above and below the zero-energy plane. For proximity-induced effective pairing $\tilde{\Delta}$ less than the amplitude of this oscillation, the system is not fully gapped. Rather, it is only gapped near the 6 band crossings of the two triangular Fermi curves (red circles in Fig. \ref{fermicontour}).

\begin{figure}
     \subfloat[]{\includegraphics[width=0.45\textwidth]{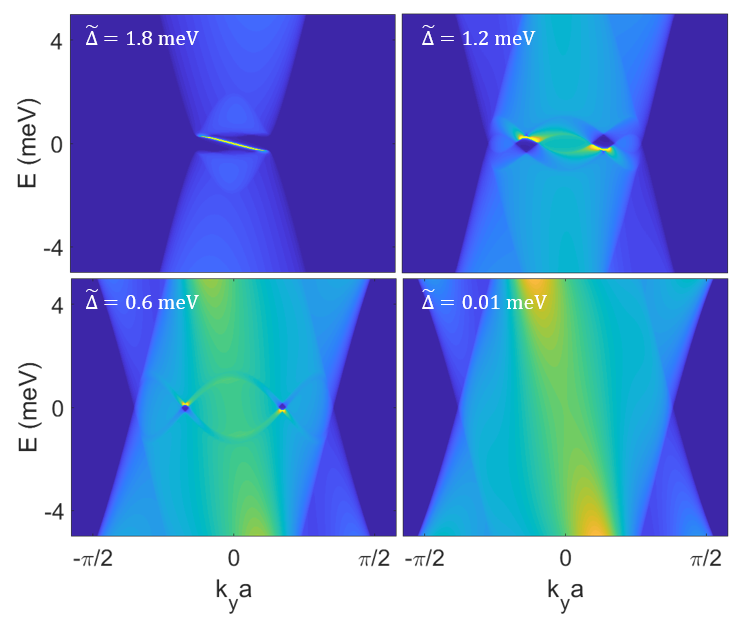}
         \label{gappedcombined}}
         
    \subfloat[]{\includegraphics[width=0.45\textwidth]{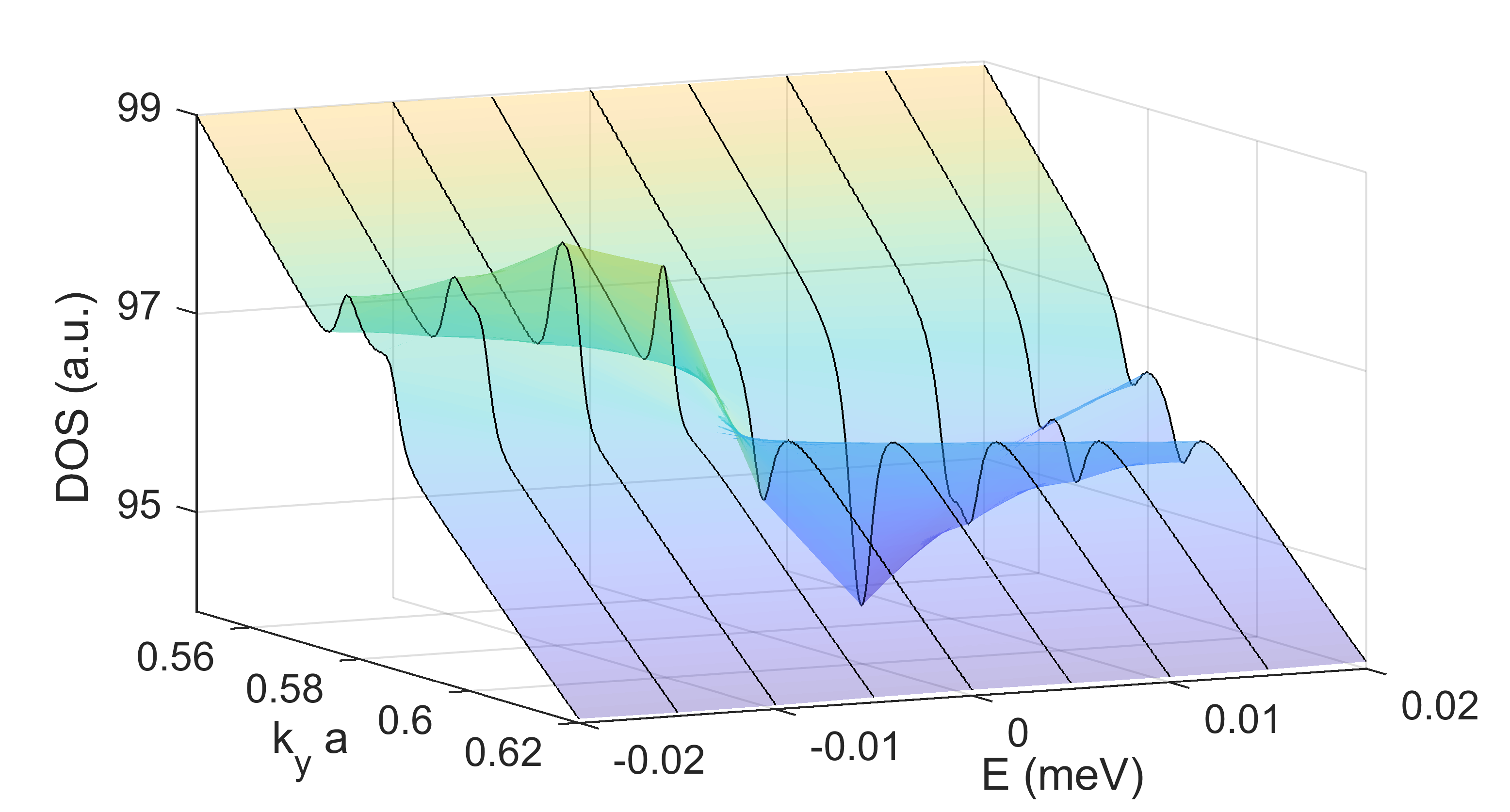}
         \label{crosssecsurf}}
         
        \caption{(a) Momentum-resolved density of states $N(k,E)=-\frac{1}{\pi}\text{Im}(G_{00}(k,E))$ (a. u.) of the interface model in a half-infinite plane, projected to its terminating zigzag edge along the $k_y$ direction. The four plots differ only in their value of the effective pairing amplitude $\tilde{\Delta}$. The top left panel shows that a chiral edge mode is present along the half-plane's edge for (unrealistically) large $\tilde{\Delta}$. As we decrease $\tilde \Delta$ and the gap closes in parts of the Fermi surface, the edge mode remains, coexisting with bulk states at the Fermi level. Since $\tilde{\Delta} = C \frac{\alpha}{J}\Delta$ (where $\Delta$ is the pairing potential of the substrate) and $ C \frac{\alpha}{J} \sim 10^{-2}$, only the bottom-right heatmap has a realistic value. Topological features are present in this case, with  $\tilde{\Delta} = 0.01$ meV, but are too weak to be visible in the heatmap. (b) Cross-sections of $N(k,E)$ in a small area of the $\tilde{\Delta} = 0.01$ meV case (bottom-right subfigure of (a)). A Fano resonance occurs for a small energy range near the Fermi energy, providing a signature of the topological state even at this realistic value of $\tilde{\Delta}$.}
        \label{gapped}
\end{figure}

Despite not being fully gapped, this system still displays topological features. We demonstrate this in Fig.~\ref{gappedcombined}, where we simulate the edge spectrum of a half-infinite plane for various values of $\tilde{\Delta}$, the effective superconducting gap width. This is related to the bare pairing term by $\tilde{\Delta} = C \frac{\alpha}{J} \Delta$ for small $\frac{\alpha}{J}$, where $C$ is a system-specific, unitless constant of order 1 (for this model, $C = 0.38$, yielding $\tilde{\Delta} = 0.0026 \Delta$).

The plots show the edge-projected density of states $N(k,E)=-\frac{1}{\pi}\text{Im}(G_{00}(k,E))$, where $k$ is momentum along the edge and $G_{00}$ is the edge-projected Green's function derived using the method of \cite{Sancho1985}. We start by considering a version of this model with an unrealistically large superconducting gap such that there are no bulk states at the Fermi energy (top-left of Fig. \ref{gappedcombined}). As expected from a fully-gapped chiral TSC, this system supports chiral edge states. As the gap size is decreased to realistic values, the edge states remain, though they are now accompanied by bulk states at the Fermi energy. These edge states, which can be seen in the density of states plot (Fig. \ref{gappedcombined}), appear only on the system edge, and only for nonzero $\tilde{\Delta}$.

Since the spins at opposite momenta contain only a small component ($\sim 1\%$) pointing in opposite directions, the effective superconducting gap is two orders of magnitude smaller than $\Delta$, the pair potential in the Pb substrate. For this reason, only the bottom-right plot in Fig. \ref{gappedcombined} is realistic.

However, even at realistic values of $\tilde{\Delta}$, these topological edge modes are present. Though they are not visible in the heatmap when the system is modeled realistically, they result in a slight (order 1\%) trough in the density of states in a small section of the Fermi surface (Fig.~\ref{crosssecsurf}). This signature, which arises from a Fano-like resonance between the discrete edge modes and the bulk continuum \cite{Fano1961}, is expected to be present in the Pb-CrI$_3$ interface. As mentioned in Sec. \ref{Theory}, however, since the topological states are surrounded by bulk states in momentum space, they are likely to be very susceptible to disorder. For this reason, the predicted resonance may demand a very clean system.

For the gapped TSC phase to exist in a hybrid system, its effective pairing would need to be comparable to the energy scale of its in-plane $C_2$-breaking terms. However, the former is a superconducting energy, on the order of $k_B T_c \sim 10^{-1} - 10^0$ meV (where $k_B$ is Boltzmann's constant and $T_c$ is the critical temperature of the superconductor). Meanwhile, the latter is a band structure feature, and as such is typically on the order of $10^1 - 10^3$ meV. Thus, topological phases in HM-SC interfaces will typically be gapless unless the HM is symmetric under in-plane $C_2$, forcing the latter terms to vanish.

\section{Conclusion}
\label{Conclusion}

We have demonstrated the existence of a gapless topological superconducting phase in a Pb-CrI$_3$ interface, and, more broadly, in similar 2D systems that do not have an in-plane $C_2$ symmetry. As a consequence, our finding identifies a significant restriction on the magnetic monolayers that may host a fully-gapped TSC where topologically-protected chiral Majorana modes may be measured. Very few magnetic monolayers that do not break $C_2$ symmetry have been found as of yet; thus, this study motivates the search for 2D magnetic materials with this symmetry, such as rectangular-lattice monolayers.

In addition, the possibility of inducing a ferromagnetic insulator to become a half-metal via charge transfer vastly expands the number of potential hybrid TSC candidate materials. Insulators with spin-polarized bands represent a promising direction in realizing topological superconductivity, both gapped and gapless.

\section*{Acknowledgements}

We acknowledge Tobias Holder and Huixia Fu for helpful conversations and advice. B.Y. acknowledges financial support by the Willner Family Leadership Institute for the Weizmann Institute of Science, the Benoziyo Endowment Fund for the Advancement of Science, the Ruth and Herman Albert Scholars Program for New Scientists, and the Israel Science Foundation (ISF 1251/19). This work was partially supported by the European Union’s Horizon 2020 research and innovation programme (Grant Agreement LEGOTOP No. 788715), the DFG (CRC/Transregio 183, EI 519/7-1), and ISF MAFAT Quantum Science and Technology (2074/19).

\appendix
\section{Tight-Binding Model}
\label{tb}

Here, we describe our tight-binding model of the Pb-CrI$_3$ interface. This does not include superconductivity, which is added afterward via the BdG formalism as outlined in Sec.~\ref{SC}. The model is closely analogous to the standard tight-binding model for graphene, which shares this system's honeycomb lattice structure.

\begin{figure}
	\centering
      \includegraphics[width=0.3\textwidth]{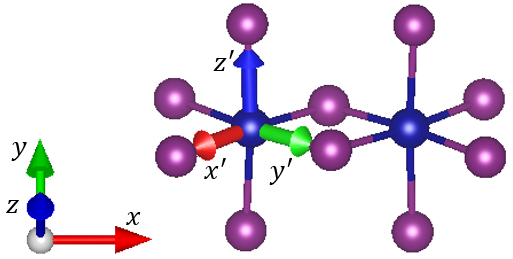}
      \caption{Iodine basis (primed coordinates) relative to lab basis. Orbitals are with respect to the iodine basis, so the relevant orbitals for this system are $d_{(x^\prime)^2 - (y^\prime)^2}$ and $d_{(z^\prime)^2}$. Though the iodine atoms do not form perfectly orthogonal axes after ionic relaxation, this effect is not large enough to be relevant to the model.}
        \label{orientation}
\end{figure}

The bands of the Pb-CrI$_3$ interface near the Fermi level are well-described by 3 binary degrees of freedom. The honeycomb lattice consists of 2 sublattices, denoted $A$ and $B$, the states can have spin up or spin down, and finally there is an orbital degree of freedom consisting of 2 degenerate orbitals, $d_{x^2 - y^2}$ and $d_{z^2}$; the remaining $d$-orbitals are gapped out by crystal field splitting. These orbitals are defined with respect to a basis formed by vectors that extend from chromium atoms to adjacent iodine atoms (see Fig. \ref{orientation}), but all other coordinates mentioned here are in the lab basis.

The most general-possible tight-binding Hamiltonian up to nearest-neighbor interactions has the form

\begin{eqnarray}\label{ham_eq}
H &=& H_{\text{OS}} + \sum_{\alpha = 1}^{3} e^{i \lambda_3 (\bm{\delta}_\alpha \cdot \bm{k})} T^\alpha .
\end{eqnarray}
Here, $H_{\text{OS}}$ represents on-site terms, while each hopping matrix $T^\alpha$ corresponds to one of the momentum vectors $\bm{\delta}_\alpha$ that takes a state on an $A$ site to an adjacent $B$ site. The Pauli z-matrix $\lambda_3$ in the sublattice degree of freedom is required to preserve Hermiticity. Generally, the $8 \cross 8$ Hermitian hopping matrix $T^\alpha$ has the form

\begin{eqnarray}\label{T_eq}
T^\alpha &=& \sum_{i,j,k} t^\alpha_{ijk} (\lambda_i \sigma_j \omega_k),
\end{eqnarray}
where the lattice, spin, and orbital degrees of freedom are acted upon by the matrices $\lambda_i$, $\sigma_j$, and $\omega_k$, respectively. The symbols $\lambda_0$, $\sigma_0$, and $\omega_0$ denote $2 \cross 2$ identity matrices in these degrees of freedom, while indices of 1 through 3 each denote the corresponding Pauli matrices.

Before the CrI$_3$ is brought into contact with the Pb, breaking $M_z$ (mirror symmetry along the direction perpendicular to the plane), the system has 3 discrete symmetries, each of which can be written as a matrix operator (times the complex conjugation operator $K$ for the case of anti-unitary symmetries):

\begin{eqnarray}\label{sym_eq}
M_x T &=& -\lambda_1 \sigma_3 \omega_3 K \nonumber \\
M_y M_z T &=& -\lambda_0 \sigma_3 \omega_3 K \\
C_3 &=& \lambda_0 (\frac{1}{2} \sigma_0 + i\frac{\sqrt{3}}{2} \sigma_3) (\frac{1}{2} \omega_0 + i\frac{\sqrt{3}}{2}\omega_{2}). \nonumber
\end{eqnarray}
Here, $T$ is time-reversal, $C_3$ is a 120-degree counterclockwise rotation about a hexagon center, and $M_i$ is a mirror symmetry in the $x_i$ direction about a hexagon center. Note that the product of the first two symmetries is $M_x M_y M_z = P$, confirming that isolated CrI$_3$ is inversion-symmetric.

Since $C_3 T^1 C_3^{\dagger} = T^2$ and $C_3 T^2 C_3^{\dagger} = T^3$, there is a relation between the coefficients $t^\alpha_{ijk}$, allowing us to express the full Hamiltonian in terms of $t^1_{ijk}$. Of these, we set to zero all the terms that change sign under $M_x T$ and $M_y M_z T$, so that the symmetries in Eq.~(\ref{sym_eq}) are conserved. This reduces the set of nonzero parameters to 10:

\begin{eqnarray}\label{myzt_eq}
S = \{t^1_{100} , t^1_{102} , t^1_{103} , t^1_{111} , t^1_{120} , \\ t^1_{122} , t^1_{123} , t^1_{130} , t^1_{132} , t^1_{133} \}. \nonumber
\end{eqnarray}

If the Pb substrate is present, then the $M_y M_z T$ symmetry is broken \footnote{Technically, $M_x T$ is also broken, since the second layer of Pb from the surface in the [111] orientation breaks $M_x$. However, since the symmetry is present in the topmost layer of Pb atoms where the majority of hybridization occurs, the surface states are only negligibly affected. This can be seen from the fact that Fig. \ref{interfacefermi} appears to be symmetric under $k_y \rightarrow -k_y$, which is enforced by $M_x T$.}, and fewer coefficients are forced to vanish. This contributes 10 additional terms:

\begin{eqnarray}\label{myzt_eq}
S^\prime = S + \{t^1_{200} , t^1_{202} , t^1_{203} , t^1_{211} , t^1_{220} , \\ t^1_{222} , t^1_{223} , t^1_{230} , t^1_{232} , t^1_{233} \}. \nonumber
\end{eqnarray}

Thus, the most general Hamiltonian that respects the symmetries of the system is

\begin{eqnarray}\label{sym_ham_eq}
H &=& H_{\text{OS}} + \sum_{\alpha = 1}^{3} e^{ i \lambda_3 (\bm{\delta}_\alpha \cdot \bm{k})} T^\alpha \nonumber \\
H_{\text{OS}} &=& \mu (\lambda_0 \sigma_0 \omega_0) + J (\lambda_0 \sigma_3 \omega_0) \nonumber \\
T^1 &=& \sum_{i,j,k \in S^\prime} t^1_{ijk} (\lambda_i \sigma_j \omega_k) \\
T^2 &=& C_3 T^1 C_3^\dagger \nonumber \\
T^3 &=& C_3^\dagger T^1 C_3 \nonumber.
\end{eqnarray}

For ease of interpretation, rather than performing a fit across the full parameter space, we choose a minimal number of terms to reproduce qualitative features of the band structure, and set the rest to 0. The coefficients $t^1_{100}$, $t^1_{102}$, $t^1_{103}$, $t^1_{130}$, and $t^1_{133}$ contribute the inversion-preserving elements of the band shape, since they are invariant under $M_yM_zT$, while the single coefficient $t^1_{202}$ breaks inversion, reducing the rotational symmetry from 6-fold to 3-fold. The inversion-symmetric spin coefficient $t^1_{111}$ determines the magnitude of the in-plane spin texture that winds twice around the Fermi surface. Similarly, $t^1_{211}$ determines the magnitude of its inversion-breaking equivalent, the in-plane Rashba texture.

The values of the nearest-neighbor hopping terms, as well as the on-site chemical potential $\mu$ and exchange field $J$, were determined by qualitatively matching the band structure to that given by DFT. The on-site pairing amplitude $\Delta$ was assumed to be of the same order as that of isolated Pb, which is $\sim1$ meV.

The spin-texture terms ($t^1_{111}$ and $t^1_{211}$) are proportional to the in-plane fraction of the spin ($\sim1\%$) times $J$ (735 meV). For $t^1_{111}$, we take the average component of the in-plane spin which is parallel to its opposite-momentum state, and $t^1_{211}$ is in principle calculated using the in-plane component perpendicular to its opposite-momentum state.

However, this underestimates the value of $t^1_{211}$, since the states at the Fermi surface lie far from any Pb band in our finite DFT simulation, artificially suppressing terms induced by coupling with the Pb \footnote{Pb bands can be made to cross CrI$_3$ bands at the Fermi surface by changing the number of Pb layers in the simulation. However, this distorts the Fermi surface and makes the CrI$_3$ band difficult to resolve}. We instead choose a value of $t^1_{211}$ of the same order as $t^1_{111}$. To put this value in experimental context, the proximity Rashba energy between a Pb substrate and a chain of Fe atoms is believed to be approximately 50 meV \cite{Yazdani2014}. Our value (5 meV) is one order of magnitude smaller, which is reasonable since Fe atoms couple farther from the substrate surface than CrI$_3$.

Values given to the nonzero parameters in this work are summarized in Table \ref{tab:params}.

\begin{table}[b]
\caption{\label{tab:params}%
Parameters used in the simplified tight-binding model. All parameters in $H$ not listed here were set to zero. The conceptual advantage of this model is demonstrated by the fact that, apart from the inversion-preserving hopping parameters that determine the shape of the relevant band, each qualitative feature of the band structure is described by only 1 term.
}
\begin{ruledtabular}
\begin{tabular}{lcr}
\textrm{Parameter Description} &
\textrm{Symbol} &
\textrm{Value (meV)}\\
\colrule
 & $t^1_{100}$ & 10\\
 & $t^1_{102}$ & -10\\
Inversion-Preserving Hopping & $t^1_{103}$ & 40\\
 & $t^1_{130}$ & -20\\
 & $t^1_{133}$ & 10\\ \cline{2-3}
Inversion-Breaking Hopping & $t^1_{202}$ & 20\\ \cline{2-3}
Inversion-Preserving Spin Texture & $t^1_{111}$ & -5\\ \cline{2-3}
Rashba Spin Texture & $t^1_{211}$ & 5\\ \cline{2-3}
Chemical Potential & $\mu$ & 845 \\ \cline{2-3}
Exchange & $J$ & 735 \\ \cline{2-3}
Pairing (added in Sec. \ref{SC}) & $\Delta$ & 1 \\
\end{tabular}
\end{ruledtabular}
\end{table}

\section{Model with Same-Spin Pairing}
\label{triplet}

In this Appendix, we consider the effect of intrinsic same-spin pairing in the SC substrate on our model. Such pairing, which couples states in the same out-of-plane spin direction rather than states of opposite spin, may arise at the Pb substrate surface as part of a triplet superconducting term~\cite{RashbaGorkov2001,Eschrig2003}. Its presence would be significant since the relevant states in our model all have a large out-of-plane spin component in the same direction, allowing them to be paired with a strength independent of the in-plane spin rotation. Thus, it bypasses the factor of $\sim10^{-2}$ that suppresses the interface's effective pairing amplitude, discussed in the beginning of Sec.~\ref{results}.

Fermion anticommutativity requires that same-spin pairing is antisymmetric in momentum space. The simplest $C_3$-symmetric tight-binding term which accomplishes this is the next-nearest neighbor term
\begin{eqnarray}\label{sym_ham_eq}
\frac{\Delta_p}{6} \sum_{\beta = 1}^{6} [(\bm{a}_\beta)_y - i(\bm{a}_\beta)_x] e^{i(\bm{a}_\beta \cdot \bm{k})} \lambda_0 \sigma_0 \omega_0 \tau_1,
\end{eqnarray}
where the vectors $\bm{a}_\beta$ connect sites in the A (B) sublattice to all next-nearest neighboring A (B) sublattice sites. Here, $\tau_i$ is a Pauli matrix in the particle-hole Nambu space. The factor of $\frac{1}{6}$ normalizes the tunneling gap to~$\Delta_p$.

When we add this term to our Hamiltonian from Eq.~(\ref{BdG}), implicitly assuming a proximity effect, we find, as expected, that the size of the gap is of the same order as $\Delta_p$. However, this is still not sufficient to fully gap the system unless the magnitude of $\Delta_p$ is comparable to the energy scale of the $C_2$-breaking terms of $H$, the largest of which are tens of meV in magnitude. Same-spin pair potentials of this scale are highly unlikely to exist in nature, even assuming no loss of pairing strength from proximity, since they would be larger than the s-wave terms of the same substrate.

\begin{figure}
	\centering
      \includegraphics[width=0.45\textwidth]{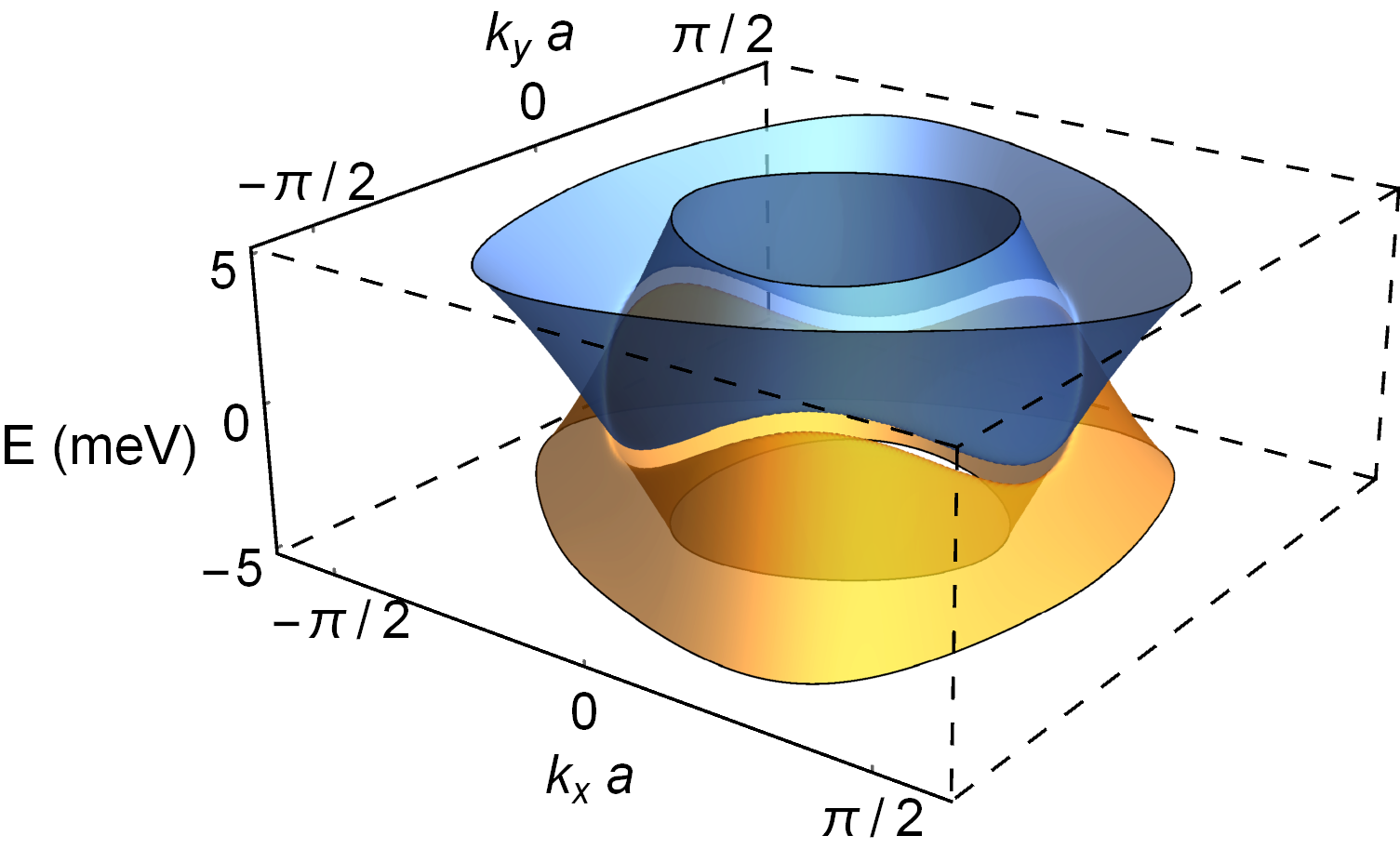}
      \caption{Band structure near the Fermi Energy (E = 0) of the modeled Pb-CrI$_3$ Hamiltonian with an added same-spin-pairing $\Delta_p$ term. Here, $\Delta_p = 1$ meV. Note that the separation between the particle and hole bands at any given momentum is significantly larger than that for a $\Delta = 1$ meV s-wave pairing interaction (Fig. \ref{fermi3d}).}
        \label{tripletGap}
\end{figure}

Figure \ref{tripletGap} shows the band structure with the addition of a same-spin pairing term $\Delta_p = 1$ meV (the s-wave pairing term $\Delta$ is set to 0, but in the case of finite $\Delta$, they would combine additively to determine the separation between the bands). In direct analogy with the results from Sec. \ref{results}, we expect that for realistic values of $\Delta_p$, we obtain a gapless topological phase with an enhancement of the resonance features near the Fermi energy. Since the effects of the $\Delta_p$ term are 2 orders of magnitude stronger than the corresponding s-wave term, even a small contribution could have relatively large effects on the system's edge states.

Thus, we conclude that while intrinsic same-spin pairing in the substrate would not affect the phase of the system, it may enable its topological features to be more readily detected.

\bibliographystyle{aipnum4-1}
\bibliography{mybib}

\begin{thebibliography}{31}%
\makeatletter
\providecommand \@ifxundefined [1]{%
 \@ifx{#1\undefined}
}%
\providecommand \@ifnum [1]{%
 \ifnum #1\expandafter \@firstoftwo
 \else \expandafter \@secondoftwo
 \fi
}%
\providecommand \@ifx [1]{%
 \ifx #1\expandafter \@firstoftwo
 \else \expandafter \@secondoftwo
 \fi
}%
\providecommand \natexlab [1]{#1}%
\providecommand \enquote  [1]{``#1''}%
\providecommand \bibnamefont  [1]{#1}%
\providecommand \bibfnamefont [1]{#1}%
\providecommand \citenamefont [1]{#1}%
\providecommand \href@noop [0]{\@secondoftwo}%
\providecommand \href [0]{\begingroup \@sanitize@url \@href}%
\providecommand \@href[1]{\@@startlink{#1}\@@href}%
\providecommand \@@href[1]{\endgroup#1\@@endlink}%
\providecommand \@sanitize@url [0]{\catcode `\\12\catcode `\$12\catcode
  `\&12\catcode `\#12\catcode `\^12\catcode `\_12\catcode `\%12\relax}%
\providecommand \@@startlink[1]{}%
\providecommand \@@endlink[0]{}%
\providecommand \url  [0]{\begingroup\@sanitize@url \@url }%
\providecommand \@url [1]{\endgroup\@href {#1}{\urlprefix }}%
\providecommand \urlprefix  [0]{URL }%
\providecommand \Eprint [0]{\href }%
\providecommand \doibase [0]{http://dx.doi.org/}%
\providecommand \selectlanguage [0]{\@gobble}%
\providecommand \bibinfo  [0]{\@secondoftwo}%
\providecommand \bibfield  [0]{\@secondoftwo}%
\providecommand \translation [1]{[#1]}%
\providecommand \BibitemOpen [0]{}%
\providecommand \bibitemStop [0]{}%
\providecommand \bibitemNoStop [0]{.\EOS\space}%
\providecommand \EOS [0]{\spacefactor3000\relax}%
\providecommand \BibitemShut  [1]{\csname bibitem#1\endcsname}%
\let\auto@bib@innerbib\@empty
\bibitem [{\citenamefont {Nayak}\ \emph {et~al.}(2008)\citenamefont {Nayak},
  \citenamefont {Simon}, \citenamefont {Stern}, \citenamefont {Freedman},\ and\
  \citenamefont {Das~Sarma}}]{Nayak2008}%
  \BibitemOpen
  \bibfield  {author} {\bibinfo {author} {\bibfnamefont {C.}~\bibnamefont
  {Nayak}}, \bibinfo {author} {\bibfnamefont {S.~H.}\ \bibnamefont {Simon}},
  \bibinfo {author} {\bibfnamefont {A.}~\bibnamefont {Stern}}, \bibinfo
  {author} {\bibfnamefont {M.}~\bibnamefont {Freedman}}, \ and\ \bibinfo
  {author} {\bibfnamefont {S.}~\bibnamefont {Das~Sarma}},\ }\href {\doibase
  10.1103/RevModPhys.80.1083} {\bibfield  {journal} {\bibinfo  {journal} {Rev.
  Mod. Phys.}\ }\textbf {\bibinfo {volume} {80}},\ \bibinfo {pages} {1083}
  (\bibinfo {year} {2008})}\BibitemShut {NoStop}%
\bibitem [{\citenamefont {Lutchyn}\ \emph {et~al.}(2018)\citenamefont
  {Lutchyn}, \citenamefont {Bakkers}, \citenamefont {Kouwenhoven},
  \citenamefont {Krogstrup}, \citenamefont {Marcus},\ and\ \citenamefont
  {Oreg}}]{Lutchyn2018}%
  \BibitemOpen
  \bibfield  {author} {\bibinfo {author} {\bibfnamefont {R.~M.}\ \bibnamefont
  {Lutchyn}}, \bibinfo {author} {\bibfnamefont {E.~P. A.~M.}\ \bibnamefont
  {Bakkers}}, \bibinfo {author} {\bibfnamefont {L.~P.}\ \bibnamefont
  {Kouwenhoven}}, \bibinfo {author} {\bibfnamefont {P.}~\bibnamefont
  {Krogstrup}}, \bibinfo {author} {\bibfnamefont {C.~M.}\ \bibnamefont
  {Marcus}}, \ and\ \bibinfo {author} {\bibfnamefont {Y.}~\bibnamefont
  {Oreg}},\ }\href {\doibase 10.1038/s41578-018-0003-1} {\bibfield  {journal}
  {\bibinfo  {journal} {Nature Reviews Materials}\ }\textbf {\bibinfo {volume}
  {3}},\ \bibinfo {pages} {52–68} (\bibinfo {year} {2018})}\BibitemShut
  {NoStop}%
\bibitem [{\citenamefont {{Nadj-Perge}}\ \emph {et~al.}(2014)\citenamefont
  {{Nadj-Perge}}, \citenamefont {{Drozdov}}, \citenamefont {{Li}},
  \citenamefont {{Chen}}, \citenamefont {{Jeon}}, \citenamefont {{Seo}},
  \citenamefont {{MacDonald}}, \citenamefont {{Bernevig}},\ and\ \citenamefont
  {{Yazdani}}}]{Yazdani2014}%
  \BibitemOpen
  \bibfield  {author} {\bibinfo {author} {\bibfnamefont {S.}~\bibnamefont
  {{Nadj-Perge}}}, \bibinfo {author} {\bibfnamefont {I.~K.}\ \bibnamefont
  {{Drozdov}}}, \bibinfo {author} {\bibfnamefont {J.}~\bibnamefont {{Li}}},
  \bibinfo {author} {\bibfnamefont {H.}~\bibnamefont {{Chen}}}, \bibinfo
  {author} {\bibfnamefont {S.}~\bibnamefont {{Jeon}}}, \bibinfo {author}
  {\bibfnamefont {J.}~\bibnamefont {{Seo}}}, \bibinfo {author} {\bibfnamefont
  {A.~H.}\ \bibnamefont {{MacDonald}}}, \bibinfo {author} {\bibfnamefont
  {B.~A.}\ \bibnamefont {{Bernevig}}}, \ and\ \bibinfo {author} {\bibfnamefont
  {A.}~\bibnamefont {{Yazdani}}},\ }\href {\doibase 10.1126/science.1259327}
  {\bibfield  {journal} {\bibinfo  {journal} {Science}\ }\textbf {\bibinfo
  {volume} {346}},\ \bibinfo {pages} {602} (\bibinfo {year}
  {2014})}\BibitemShut {NoStop}%
\bibitem [{\citenamefont {Vaitiekėnas}\ \emph {et~al.}(2018)\citenamefont
  {Vaitiekėnas}, \citenamefont {Deng}, \citenamefont {Krogstrup},\ and\
  \citenamefont {Marcus}}]{Vaitiekenas2018}%
  \BibitemOpen
  \bibfield  {author} {\bibinfo {author} {\bibfnamefont {S.}~\bibnamefont
  {Vaitiekėnas}}, \bibinfo {author} {\bibfnamefont {M.~T.}\ \bibnamefont
  {Deng}}, \bibinfo {author} {\bibfnamefont {P.}~\bibnamefont {Krogstrup}}, \
  and\ \bibinfo {author} {\bibfnamefont {C.~M.}\ \bibnamefont {Marcus}},\
  }\href@noop {} {\  (\bibinfo {year} {2018})},\ \Eprint
  {http://arxiv.org/abs/1809.05513} {arXiv:1809.05513 [cond-mat.mes-hall]}
  \BibitemShut {NoStop}%
\bibitem [{\citenamefont {Ren}\ \emph {et~al.}(2019)\citenamefont {Ren},
  \citenamefont {Pientka}, \citenamefont {Hart}, \citenamefont {Pierce},
  \citenamefont {Kosowsky}, \citenamefont {Lunczer}, \citenamefont {Schlereth},
  \citenamefont {Scharf}, \citenamefont {Hankiewicz}, \citenamefont
  {Molenkamp}, \citenamefont {Halperin},\ and\ \citenamefont
  {Yacoby}}]{Yacoby2019}%
  \BibitemOpen
  \bibfield  {author} {\bibinfo {author} {\bibfnamefont {H.}~\bibnamefont
  {Ren}}, \bibinfo {author} {\bibfnamefont {F.}~\bibnamefont {Pientka}},
  \bibinfo {author} {\bibfnamefont {S.}~\bibnamefont {Hart}}, \bibinfo {author}
  {\bibfnamefont {A.~T.}\ \bibnamefont {Pierce}}, \bibinfo {author}
  {\bibfnamefont {M.}~\bibnamefont {Kosowsky}}, \bibinfo {author}
  {\bibfnamefont {L.}~\bibnamefont {Lunczer}}, \bibinfo {author} {\bibfnamefont
  {R.}~\bibnamefont {Schlereth}}, \bibinfo {author} {\bibfnamefont
  {B.}~\bibnamefont {Scharf}}, \bibinfo {author} {\bibfnamefont {E.~M.}\
  \bibnamefont {Hankiewicz}}, \bibinfo {author} {\bibfnamefont {L.~W.}\
  \bibnamefont {Molenkamp}}, \bibinfo {author} {\bibfnamefont {B.~I.}\
  \bibnamefont {Halperin}}, \ and\ \bibinfo {author} {\bibfnamefont
  {A.}~\bibnamefont {Yacoby}},\ }\href
  {https://doi.org/10.1038/s41586-019-1148-9} {\bibfield  {journal} {\bibinfo
  {journal} {Nature}\ }\textbf {\bibinfo {volume} {569}},\ \bibinfo {pages}
  {93} (\bibinfo {year} {2019})}\BibitemShut {NoStop}%
\bibitem [{\citenamefont {Fornieri}\ \emph {et~al.}(2019)\citenamefont
  {Fornieri}, \citenamefont {Whiticar}, \citenamefont {Setiawan}, \citenamefont
  {Portolés}, \citenamefont {Drachmann}, \citenamefont {Keselman},
  \citenamefont {Gronin}, \citenamefont {Thomas}, \citenamefont {Wang},
  \citenamefont {Kallaher}, \citenamefont {Gardner}, \citenamefont {Berg},
  \citenamefont {Manfra}, \citenamefont {Stern}, \citenamefont {Marcus},\ and\
  \citenamefont {Nichele}}]{Fornieri2019}%
  \BibitemOpen
  \bibfield  {author} {\bibinfo {author} {\bibfnamefont {A.}~\bibnamefont
  {Fornieri}}, \bibinfo {author} {\bibfnamefont {A.~M.}\ \bibnamefont
  {Whiticar}}, \bibinfo {author} {\bibfnamefont {F.}~\bibnamefont {Setiawan}},
  \bibinfo {author} {\bibfnamefont {E.}~\bibnamefont {Portolés}}, \bibinfo
  {author} {\bibfnamefont {A.~C.~C.}\ \bibnamefont {Drachmann}}, \bibinfo
  {author} {\bibfnamefont {A.}~\bibnamefont {Keselman}}, \bibinfo {author}
  {\bibfnamefont {S.}~\bibnamefont {Gronin}}, \bibinfo {author} {\bibfnamefont
  {C.}~\bibnamefont {Thomas}}, \bibinfo {author} {\bibfnamefont
  {T.}~\bibnamefont {Wang}}, \bibinfo {author} {\bibfnamefont {R.}~\bibnamefont
  {Kallaher}}, \bibinfo {author} {\bibfnamefont {G.~C.}\ \bibnamefont
  {Gardner}}, \bibinfo {author} {\bibfnamefont {E.}~\bibnamefont {Berg}},
  \bibinfo {author} {\bibfnamefont {M.~J.}\ \bibnamefont {Manfra}}, \bibinfo
  {author} {\bibfnamefont {A.}~\bibnamefont {Stern}}, \bibinfo {author}
  {\bibfnamefont {C.~M.}\ \bibnamefont {Marcus}}, \ and\ \bibinfo {author}
  {\bibfnamefont {F.}~\bibnamefont {Nichele}},\ }\href
  {https://doi.org/10.1038/s41586-019-1068-8} {\bibfield  {journal} {\bibinfo
  {journal} {Nature}\ }\textbf {\bibinfo {volume} {569}},\ \bibinfo {pages}
  {89} (\bibinfo {year} {2019})}\BibitemShut {NoStop}%
\bibitem [{\citenamefont {Miró}, \citenamefont {Audiffred},\ and\
  \citenamefont {Heine}(2014)}]{Heine2014}%
  \BibitemOpen
  \bibfield  {author} {\bibinfo {author} {\bibfnamefont {P.}~\bibnamefont
  {Miró}}, \bibinfo {author} {\bibfnamefont {M.}~\bibnamefont {Audiffred}}, \
  and\ \bibinfo {author} {\bibfnamefont {T.}~\bibnamefont {Heine}},\ }\href
  {\doibase 10.1039/C4CS00102H} {\bibfield  {journal} {\bibinfo  {journal}
  {Chem. Soc. Rev.}\ }\textbf {\bibinfo {volume} {43}},\ \bibinfo {pages}
  {6537} (\bibinfo {year} {2014})}\BibitemShut {NoStop}%
\bibitem [{\citenamefont {Duckheim}\ and\ \citenamefont
  {Brouwer}(2011)}]{Brouwer2011}%
  \BibitemOpen
  \bibfield  {author} {\bibinfo {author} {\bibfnamefont {M.}~\bibnamefont
  {Duckheim}}\ and\ \bibinfo {author} {\bibfnamefont {P.~W.}\ \bibnamefont
  {Brouwer}},\ }\href {\doibase 10.1103/PhysRevB.83.054513} {\bibfield
  {journal} {\bibinfo  {journal} {Phys. Rev. B}\ }\textbf {\bibinfo {volume}
  {83}},\ \bibinfo {pages} {054513} (\bibinfo {year} {2011})}\BibitemShut
  {NoStop}%
\bibitem [{\citenamefont {Sau}\ \emph {et~al.}(2010)\citenamefont {Sau},
  \citenamefont {Tewari}, \citenamefont {Lutchyn}, \citenamefont {Stanescu},\
  and\ \citenamefont {Das~Sarma}}]{Sau2010}%
  \BibitemOpen
  \bibfield  {author} {\bibinfo {author} {\bibfnamefont {J.~D.}\ \bibnamefont
  {Sau}}, \bibinfo {author} {\bibfnamefont {S.}~\bibnamefont {Tewari}},
  \bibinfo {author} {\bibfnamefont {R.~M.}\ \bibnamefont {Lutchyn}}, \bibinfo
  {author} {\bibfnamefont {T.~D.}\ \bibnamefont {Stanescu}}, \ and\ \bibinfo
  {author} {\bibfnamefont {S.}~\bibnamefont {Das~Sarma}},\ }\href {\doibase
  10.1103/PhysRevB.82.214509} {\bibfield  {journal} {\bibinfo  {journal} {Phys.
  Rev. B}\ }\textbf {\bibinfo {volume} {82}},\ \bibinfo {pages} {214509}
  (\bibinfo {year} {2010})}\BibitemShut {NoStop}%
\bibitem [{\citenamefont {{Chung}}\ \emph {et~al.}(2011)\citenamefont
  {{Chung}}, \citenamefont {{Zhang}}, \citenamefont {{Qi}},\ and\ \citenamefont
  {{Zhang}}}]{SCZhang2011}%
  \BibitemOpen
  \bibfield  {author} {\bibinfo {author} {\bibfnamefont {S.~B.}\ \bibnamefont
  {{Chung}}}, \bibinfo {author} {\bibfnamefont {H.-J.}\ \bibnamefont
  {{Zhang}}}, \bibinfo {author} {\bibfnamefont {X.-L.}\ \bibnamefont {{Qi}}}, \
  and\ \bibinfo {author} {\bibfnamefont {S.-C.}\ \bibnamefont {{Zhang}}},\
  }\href {\doibase 10.1103/PhysRevB.84.060510} {\bibfield  {journal} {\bibinfo
  {journal} {Physical Review B}\ }\textbf {\bibinfo {volume} {84}},\ \bibinfo
  {pages} {060510} (\bibinfo {year} {2011})}\BibitemShut {NoStop}%
\bibitem [{\citenamefont {Hao}\ and\ \citenamefont {Ting}(2016)}]{Hao2016}%
  \BibitemOpen
  \bibfield  {author} {\bibinfo {author} {\bibfnamefont {L.}~\bibnamefont
  {Hao}}\ and\ \bibinfo {author} {\bibfnamefont {C.~S.}\ \bibnamefont {Ting}},\
  }\href {\doibase 10.1103/PhysRevB.94.134513} {\bibfield  {journal} {\bibinfo
  {journal} {Phys. Rev. B}\ }\textbf {\bibinfo {volume} {94}},\ \bibinfo
  {pages} {134513} (\bibinfo {year} {2016})}\BibitemShut {NoStop}%
\bibitem [{\citenamefont {Palacio-Morales}\ \emph {et~al.}(2019)\citenamefont
  {Palacio-Morales}, \citenamefont {Mascot}, \citenamefont {Cocklin},
  \citenamefont {Kim}, \citenamefont {Rachel}, \citenamefont {Morr},\ and\
  \citenamefont {Wiesendanger}}]{Wiesendanger2019}%
  \BibitemOpen
  \bibfield  {author} {\bibinfo {author} {\bibfnamefont {A.}~\bibnamefont
  {Palacio-Morales}}, \bibinfo {author} {\bibfnamefont {E.}~\bibnamefont
  {Mascot}}, \bibinfo {author} {\bibfnamefont {S.}~\bibnamefont {Cocklin}},
  \bibinfo {author} {\bibfnamefont {H.}~\bibnamefont {Kim}}, \bibinfo {author}
  {\bibfnamefont {S.}~\bibnamefont {Rachel}}, \bibinfo {author} {\bibfnamefont
  {D.~K.}\ \bibnamefont {Morr}}, \ and\ \bibinfo {author} {\bibfnamefont
  {R.}~\bibnamefont {Wiesendanger}},\ }\href
  {https://advances.sciencemag.org/content/5/7/eaav6600} {\bibfield  {journal}
  {\bibinfo  {journal} {Science Advances}\ }\textbf {\bibinfo {volume} {5}}
  (\bibinfo {year} {2019})}\BibitemShut {NoStop}%
\bibitem [{\citenamefont {Baral}\ \emph {et~al.}(2019)\citenamefont {Baral},
  \citenamefont {Fu}, \citenamefont {Zadorozhnyi}, \citenamefont {Dulal},
  \citenamefont {Wang}, \citenamefont {Shrestha}, \citenamefont {Erugu},
  \citenamefont {Tang}, \citenamefont {Dahnovsky}, \citenamefont {Tian},\ and\
  \citenamefont {Chien}}]{Baral2019}%
  \BibitemOpen
  \bibfield  {author} {\bibinfo {author} {\bibfnamefont {D.}~\bibnamefont
  {Baral}}, \bibinfo {author} {\bibfnamefont {Z.}~\bibnamefont {Fu}}, \bibinfo
  {author} {\bibfnamefont {A.~S.}\ \bibnamefont {Zadorozhnyi}}, \bibinfo
  {author} {\bibfnamefont {R.}~\bibnamefont {Dulal}}, \bibinfo {author}
  {\bibfnamefont {A.}~\bibnamefont {Wang}}, \bibinfo {author} {\bibfnamefont
  {N.}~\bibnamefont {Shrestha}}, \bibinfo {author} {\bibfnamefont
  {U.}~\bibnamefont {Erugu}}, \bibinfo {author} {\bibfnamefont
  {J.}~\bibnamefont {Tang}}, \bibinfo {author} {\bibfnamefont {Y.}~\bibnamefont
  {Dahnovsky}}, \bibinfo {author} {\bibfnamefont {J.}~\bibnamefont {Tian}}, \
  and\ \bibinfo {author} {\bibfnamefont {T.}~\bibnamefont {Chien}},\
  }\href@noop {} {} (\bibinfo {year} {2019}),\ \Eprint
  {http://arxiv.org/abs/1909.00074} {arXiv:1909.00074 [cond-mat.mtrl-sci]}
  \BibitemShut {NoStop}%
\bibitem [{\citenamefont {Chen}\ \emph {et~al.}(2019)\citenamefont {Chen},
  \citenamefont {Sun}, \citenamefont {Wang}, \citenamefont {Gu}, \citenamefont
  {Xu}, \citenamefont {Wu},\ and\ \citenamefont {Gao}}]{Chen2019}%
  \BibitemOpen
  \bibfield  {author} {\bibinfo {author} {\bibfnamefont {W.}~\bibnamefont
  {Chen}}, \bibinfo {author} {\bibfnamefont {Z.}~\bibnamefont {Sun}}, \bibinfo
  {author} {\bibfnamefont {Z.}~\bibnamefont {Wang}}, \bibinfo {author}
  {\bibfnamefont {L.}~\bibnamefont {Gu}}, \bibinfo {author} {\bibfnamefont
  {X.}~\bibnamefont {Xu}}, \bibinfo {author} {\bibfnamefont {S.}~\bibnamefont
  {Wu}}, \ and\ \bibinfo {author} {\bibfnamefont {C.}~\bibnamefont {Gao}},\
  }\href {\doibase 10.1126/science.aav1937} {\bibfield  {journal} {\bibinfo
  {journal} {Science}\ }\textbf {\bibinfo {volume} {366}},\ \bibinfo {pages}
  {983} (\bibinfo {year} {2019})}\BibitemShut {NoStop}%
\bibitem [{\citenamefont {Li}\ \emph {et~al.}(2019)\citenamefont {Li},
  \citenamefont {Wang}, \citenamefont {Zhang}, \citenamefont {Chen},
  \citenamefont {Guo}, \citenamefont {Ji},\ and\ \citenamefont
  {Zhong}}]{Li2019}%
  \BibitemOpen
  \bibfield  {author} {\bibinfo {author} {\bibfnamefont {P.}~\bibnamefont
  {Li}}, \bibinfo {author} {\bibfnamefont {C.}~\bibnamefont {Wang}}, \bibinfo
  {author} {\bibfnamefont {J.}~\bibnamefont {Zhang}}, \bibinfo {author}
  {\bibfnamefont {S.}~\bibnamefont {Chen}}, \bibinfo {author} {\bibfnamefont
  {D.}~\bibnamefont {Guo}}, \bibinfo {author} {\bibfnamefont {W.}~\bibnamefont
  {Ji}}, \ and\ \bibinfo {author} {\bibfnamefont {D.}~\bibnamefont {Zhong}},\
  }\href@noop {} {\enquote {\bibinfo {title} {Single-layer cri3 grown by
  molecular beam epitaxy},}\ } (\bibinfo {year} {2019}),\ \Eprint
  {http://arxiv.org/abs/1912.02559} {arXiv:1912.02559 [cond-mat.mtrl-sci]}
  \BibitemShut {NoStop}%
\bibitem [{\citenamefont {{Huang}}\ \emph {et~al.}(2017)\citenamefont
  {{Huang}}, \citenamefont {{Clark}}, \citenamefont {{Navarro-Moratalla}},
  \citenamefont {{Klein}}, \citenamefont {{Cheng}}, \citenamefont {{Seyler}},
  \citenamefont {{Zhong}}, \citenamefont {{Schmidgall}}, \citenamefont
  {{McGuire}}, \citenamefont {{Cobden}}, \citenamefont {{Yao}}, \citenamefont
  {{Xiao}}, \citenamefont {{Jarillo-Herrero}},\ and\ \citenamefont
  {{Xu}}}]{Huang2017}%
  \BibitemOpen
  \bibfield  {author} {\bibinfo {author} {\bibfnamefont {B.}~\bibnamefont
  {{Huang}}}, \bibinfo {author} {\bibfnamefont {G.}~\bibnamefont {{Clark}}},
  \bibinfo {author} {\bibfnamefont {E.}~\bibnamefont {{Navarro-Moratalla}}},
  \bibinfo {author} {\bibfnamefont {D.~R.}\ \bibnamefont {{Klein}}}, \bibinfo
  {author} {\bibfnamefont {R.}~\bibnamefont {{Cheng}}}, \bibinfo {author}
  {\bibfnamefont {K.~L.}\ \bibnamefont {{Seyler}}}, \bibinfo {author}
  {\bibfnamefont {D.}~\bibnamefont {{Zhong}}}, \bibinfo {author} {\bibfnamefont
  {E.}~\bibnamefont {{Schmidgall}}}, \bibinfo {author} {\bibfnamefont {M.~A.}\
  \bibnamefont {{McGuire}}}, \bibinfo {author} {\bibfnamefont {D.~H.}\
  \bibnamefont {{Cobden}}}, \bibinfo {author} {\bibfnamefont {W.}~\bibnamefont
  {{Yao}}}, \bibinfo {author} {\bibfnamefont {D.}~\bibnamefont {{Xiao}}},
  \bibinfo {author} {\bibfnamefont {P.}~\bibnamefont {{Jarillo-Herrero}}}, \
  and\ \bibinfo {author} {\bibfnamefont {X.}~\bibnamefont {{Xu}}},\ }\href
  {\doibase 10.1038/nature22391} {\bibfield  {journal} {\bibinfo  {journal}
  {Nature}\ }\textbf {\bibinfo {volume} {546}},\ \bibinfo {pages} {270}
  (\bibinfo {year} {2017})}\BibitemShut {NoStop}%
\bibitem [{\citenamefont {{Sato}}(2010)}]{Sato2010}%
  \BibitemOpen
  \bibfield  {author} {\bibinfo {author} {\bibfnamefont {M.}~\bibnamefont
  {{Sato}}},\ }\href {\doibase 10.1103/PhysRevB.81.220504} {\bibfield
  {journal} {\bibinfo  {journal} {\prb}\ }\textbf {\bibinfo {volume} {81}},\
  \bibinfo {eid} {220504} (\bibinfo {year} {2010})}\BibitemShut {NoStop}%
\bibitem [{\citenamefont {Wong}\ \emph {et~al.}(2013)\citenamefont {Wong},
  \citenamefont {Liu}, \citenamefont {Law},\ and\ \citenamefont
  {Lee}}]{Wong2013}%
  \BibitemOpen
  \bibfield  {author} {\bibinfo {author} {\bibfnamefont {C.~L.~M.}\
  \bibnamefont {Wong}}, \bibinfo {author} {\bibfnamefont {J.}~\bibnamefont
  {Liu}}, \bibinfo {author} {\bibfnamefont {K.~T.}\ \bibnamefont {Law}}, \ and\
  \bibinfo {author} {\bibfnamefont {P.~A.}\ \bibnamefont {Lee}},\ }\href
  {\doibase 10.1103/PhysRevB.88.060504} {\bibfield  {journal} {\bibinfo
  {journal} {Phys. Rev. B}\ }\textbf {\bibinfo {volume} {88}},\ \bibinfo
  {pages} {060504} (\bibinfo {year} {2013})}\BibitemShut {NoStop}%
\bibitem [{\citenamefont {{Baum}}\ \emph {et~al.}(2015)\citenamefont {{Baum}},
  \citenamefont {{Posske}}, \citenamefont {{Fulga}}, \citenamefont
  {{Trauzettel}},\ and\ \citenamefont {{Stern}}}]{Baum2015}%
  \BibitemOpen
  \bibfield  {author} {\bibinfo {author} {\bibfnamefont {Y.}~\bibnamefont
  {{Baum}}}, \bibinfo {author} {\bibfnamefont {T.}~\bibnamefont {{Posske}}},
  \bibinfo {author} {\bibfnamefont {I.~C.}\ \bibnamefont {{Fulga}}}, \bibinfo
  {author} {\bibfnamefont {B.}~\bibnamefont {{Trauzettel}}}, \ and\ \bibinfo
  {author} {\bibfnamefont {A.}~\bibnamefont {{Stern}}},\ }\href {\doibase
  10.1103/PhysRevB.92.045128} {\bibfield  {journal} {\bibinfo  {journal}
  {\prb}\ }\textbf {\bibinfo {volume} {92}},\ \bibinfo {eid} {045128} (\bibinfo
  {year} {2015})}\BibitemShut {NoStop}%
\bibitem [{\citenamefont {Daido}\ and\ \citenamefont
  {Yanase}(2017)}]{Daido2017}%
  \BibitemOpen
  \bibfield  {author} {\bibinfo {author} {\bibfnamefont {A.}~\bibnamefont
  {Daido}}\ and\ \bibinfo {author} {\bibfnamefont {Y.}~\bibnamefont {Yanase}},\
  }\href {\doibase 10.1103/PhysRevB.95.134507} {\bibfield  {journal} {\bibinfo
  {journal} {Phys. Rev. B}\ }\textbf {\bibinfo {volume} {95}},\ \bibinfo
  {pages} {134507} (\bibinfo {year} {2017})}\BibitemShut {NoStop}%
\bibitem [{\citenamefont {{Ying}}\ and\ \citenamefont
  {{Kamenev}}(2018)}]{Kamenev2018}%
  \BibitemOpen
  \bibfield  {author} {\bibinfo {author} {\bibfnamefont {X.}~\bibnamefont
  {{Ying}}}\ and\ \bibinfo {author} {\bibfnamefont {A.}~\bibnamefont
  {{Kamenev}}},\ }\href {\doibase 10.1103/PhysRevLett.121.086810} {\bibfield
  {journal} {\bibinfo  {journal} {Phys. Rev. Lett.}\ }\textbf {\bibinfo
  {volume} {121}},\ \bibinfo {pages} {086810} (\bibinfo {year}
  {2018})}\BibitemShut {NoStop}%
\bibitem [{\citenamefont {Rashba}(1960)}]{Rashba}%
  \BibitemOpen
  \bibfield  {author} {\bibinfo {author} {\bibfnamefont {E.}~\bibnamefont
  {Rashba}},\ }\href@noop {} {\bibfield  {journal} {\bibinfo  {journal} {Sov.
  Phys. Solid State}\ }\textbf {\bibinfo {volume} {2}},\ \bibinfo {pages}
  {1109} (\bibinfo {year} {1960})}\BibitemShut {NoStop}%
\bibitem [{\citenamefont {Kresse}\ and\ \citenamefont
  {Hafner}(1993)}]{Kresse1993}%
  \BibitemOpen
  \bibfield  {author} {\bibinfo {author} {\bibfnamefont {G.}~\bibnamefont
  {Kresse}}\ and\ \bibinfo {author} {\bibfnamefont {J.}~\bibnamefont
  {Hafner}},\ }\href {\doibase 10.1103/PhysRevB.47.558} {\bibfield  {journal}
  {\bibinfo  {journal} {Phys. Rev. B}\ }\textbf {\bibinfo {volume} {47}},\
  \bibinfo {pages} {558} (\bibinfo {year} {1993})}\BibitemShut {NoStop}%
\bibitem [{\citenamefont {Kresse}\ and\ \citenamefont
  {Joubert}(1999)}]{Kresse1999}%
  \BibitemOpen
  \bibfield  {author} {\bibinfo {author} {\bibfnamefont {G.}~\bibnamefont
  {Kresse}}\ and\ \bibinfo {author} {\bibfnamefont {D.}~\bibnamefont
  {Joubert}},\ }\href {\doibase 10.1103/PhysRevB.59.1758} {\bibfield  {journal}
  {\bibinfo  {journal} {Phys. Rev. B}\ }\textbf {\bibinfo {volume} {59}},\
  \bibinfo {pages} {1758} (\bibinfo {year} {1999})}\BibitemShut {NoStop}%
\bibitem [{\citenamefont {Perdew}, \citenamefont {Burke},\ and\ \citenamefont
  {Ernzerhof}(1996)}]{GGA}%
  \BibitemOpen
  \bibfield  {author} {\bibinfo {author} {\bibfnamefont {J.~P.}\ \bibnamefont
  {Perdew}}, \bibinfo {author} {\bibfnamefont {K.}~\bibnamefont {Burke}}, \
  and\ \bibinfo {author} {\bibfnamefont {M.}~\bibnamefont {Ernzerhof}},\ }\href
  {\doibase 10.1103/PhysRevLett.77.3865} {\bibfield  {journal} {\bibinfo
  {journal} {Phys. Rev. Lett.}\ }\textbf {\bibinfo {volume} {77}},\ \bibinfo
  {pages} {3865} (\bibinfo {year} {1996})}\BibitemShut {NoStop}%
\bibitem [{\citenamefont {Sancho}, \citenamefont {Sancho},\ and\ \citenamefont
  {Rubio}(1985)}]{Sancho1985}%
  \BibitemOpen
  \bibfield  {author} {\bibinfo {author} {\bibfnamefont {M.~P.~L.}\
  \bibnamefont {Sancho}}, \bibinfo {author} {\bibfnamefont {J.~M.~L.}\
  \bibnamefont {Sancho}}, \ and\ \bibinfo {author} {\bibfnamefont
  {J.}~\bibnamefont {Rubio}},\ }\href {\doibase 10.1088/0305-4608/15/4/009}
  {\bibfield  {journal} {\bibinfo  {journal} {Journal of Physics F: Metal
  Physics}\ }\textbf {\bibinfo {volume} {15}},\ \bibinfo {pages} {851}
  (\bibinfo {year} {1985})}\BibitemShut {NoStop}%
\bibitem [{\citenamefont {Fano}(1961)}]{Fano1961}%
  \BibitemOpen
  \bibfield  {author} {\bibinfo {author} {\bibfnamefont {U.}~\bibnamefont
  {Fano}},\ }\href {\doibase 10.1103/PhysRev.124.1866} {\bibfield  {journal}
  {\bibinfo  {journal} {Phys. Rev.}\ }\textbf {\bibinfo {volume} {124}},\
  \bibinfo {pages} {1866} (\bibinfo {year} {1961})}\BibitemShut {NoStop}%
\bibitem [{Note1()}]{Note1}%
  \BibitemOpen
  \bibinfo {note} {Technically, $M_x T$ is also broken, since the second layer
  of Pb from the surface in the [111] orientation breaks $M_x$. However, since
  the symmetry is present in the topmost layer of Pb atoms where the majority
  of hybridization occurs, the surface states are only negligibly affected.
  This can be seen from the fact that Fig. \ref {interfacefermi} appears to be
  symmetric under $k_y \rightarrow -k_y$, which is enforced by $M_x
  T$.}\BibitemShut {Stop}%
\bibitem [{Note2()}]{Note2}%
  \BibitemOpen
  \bibinfo {note} {Pb bands can be made to cross CrI$_3$ bands at the Fermi
  surface by changing the number of Pb layers in the simulation. However, this
  distorts the Fermi surface and makes the CrI$_3$ band difficult to
  resolve}\BibitemShut {NoStop}%
\bibitem [{\citenamefont {{Gor'kov}}\ and\ \citenamefont
  {{Rashba}}(2001)}]{RashbaGorkov2001}%
  \BibitemOpen
  \bibfield  {author} {\bibinfo {author} {\bibfnamefont {L.~P.}\ \bibnamefont
  {{Gor'kov}}}\ and\ \bibinfo {author} {\bibfnamefont {E.~I.}\ \bibnamefont
  {{Rashba}}},\ }\href {\doibase 10.1103/PhysRevLett.87.037004} {\bibfield
  {journal} {\bibinfo  {journal} {Phys. Rev. Lett.}\ }\textbf {\bibinfo
  {volume} {87}},\ \bibinfo {pages} {037004} (\bibinfo {year}
  {2001})}\BibitemShut {NoStop}%
\bibitem [{\citenamefont {{Eschrig}}\ \emph {et~al.}(2003)\citenamefont
  {{Eschrig}}, \citenamefont {{Kopu}}, \citenamefont {{Cuevas}},\ and\
  \citenamefont {{Sch{\"o}n}}}]{Eschrig2003}%
  \BibitemOpen
  \bibfield  {author} {\bibinfo {author} {\bibfnamefont {M.}~\bibnamefont
  {{Eschrig}}}, \bibinfo {author} {\bibfnamefont {J.}~\bibnamefont {{Kopu}}},
  \bibinfo {author} {\bibfnamefont {J.~C.}\ \bibnamefont {{Cuevas}}}, \ and\
  \bibinfo {author} {\bibfnamefont {G.}~\bibnamefont {{Sch{\"o}n}}},\ }\href
  {\doibase 10.1103/PhysRevLett.90.137003} {\bibfield  {journal} {\bibinfo
  {journal} {Phys. Rev. Lett.}\ }\textbf {\bibinfo {volume} {90}},\ \bibinfo
  {pages} {137003} (\bibinfo {year} {2003})}\BibitemShut {NoStop}%
\end{thebibliography}%

\end{document}